\documentclass{emulateapj}
\usepackage{lscape}
\usepackage{longtable}

\def\sun{\hbox{$\odot$}}

\def\farcs{\hbox{$.\!\!^{\prime\prime}$}}

\slugcomment{Accepted by the Astronomical Journal.}

\shorttitle{GALEX UV Atlas of Interacting Galaxies}
\shortauthors{Smith et al.}

\begin{document}


\title{Spirals, Bridges, and Tails:
A GALEX UV Atlas of Interacting Galaxies}


\author{Beverly J. Smith}
\affil{Department of Physics and Astronomy, East Tennessee
State University, Johnson City TN  37614}
\email{smithbj@etsu.edu}

\author{Mark L. Giroux}
\affil{Department of Physics and Astronomy, East Tennessee
State University, Johnson City TN  37614}
\email{girouxm@etsu.edu}

\author{Curtis Struck}
\affil{Department of Physics and Astronomy, Iowa State University, Ames IA  50011}
\email{curt@iastate.edu}

\author{Mark Hancock}
\affil{Department of Physics, University of California Riverside,
Riverside CA 92521}
\email{mhancock@ucr.edu}

\author{Sabrina Hurlock}
\affil{Department of Physics and Astronomy, East Tennessee
State University, Johnson City TN  37614}
\email{zshh7@goldmail.etsu.edu}



\begin{abstract}

We have used the GALEX ultraviolet telescope to study stellar populations and 
star formation morphology in a well-defined sample of 42 nearby 
optically-selected 
pre-merger 
interacting galaxy pairs. 
Galaxy interactions were likely far more common in the 
early Universe than in the present, thus our study 
provides a nearby well-resolved comparison sample 
for high redshift studies.  
We have combined the GALEX NUV and FUV 
images with broadband optical maps from the Sloan Digitized Sky
Survey to investigate
the
ages and extinctions of the tidal features
and the disks.  
The distributions of the UV/optical colors of the tidal features
and the main disks of the galaxies are similar,
however, the tidal features
are bluer on average in NUV $-$ g when 
compared with their own parent
disks, thus tails and bridges are often more prominent relative
to the disks in UV images
compared to optical maps.
This effect is likely due to enhanced star formation in the
tidal features compared to the disks rather than reduced extinction, 
however, lower metallicities may also play a role.
We have identified a few new
candidate tidal dwarf galaxies in this sample.  Other interesting morphologies
such as accretion tails and `beads on a string' are also 
seen in these images.
We also identify a possible `Taffy' galaxy in our sample,
which may have been produced by a head-on collision between
two galaxies.
In only a few cases are strong tidal features seen in HI maps
but not in GALEX.

\end{abstract}



\keywords{galaxies: starbursts ---
galaxies: interactions--- 
galaxies: ultraviolet}

\section{Introduction}

Tidal disturbances have played an 
important role in reshaping galaxies and triggering star formation
over
cosmic time (see \citealp{struck99} for review).
H$\alpha$, far-infrared,
and mid-infrared observations show that 
the mass-normalized star formation rates of 
pre-merger interacting systems are 
enhanced by a factor of two on average
compared to normal spirals
(e.g., \citealp{bushouse87, kennicutt87, bushouse88,
barton00, barton03, smith07, lin07}).  Further, closer pairs
have more enhanced star formation than
wider pairs, but with
significant scatter \citep{barton00, lambas03, nikolic04, lin07}.
This effect is more difficult to see in broadband optical colors;
\citet{larson78}
found a larger scatter in the broadband
optical colors of Arp Atlas
galaxies than isolated galaxies, while
\citet{bergvall03}
found little color difference between an interacting
and a more isolated sample.

Within interacting galaxies,
star formation is often enhanced in the nuclear regions compared
to the disks \citep{hummel90, nikolic04}.
Luminous star forming regions are sometimes seen in tidal features 
\citep{schweizer78, mirabel91, mirabel92, hibbard96}.
For a sample of 25 Arp Atlas galaxies, 
\citet{schombert90} found that the B $-$ V colors of the tidal
features were somewhat bluer on average than the parent disks, 
but with significant scatter.
They conclude that this difference is due to either
enhanced star formation or lower extinction in tidal features.

With the advent of the Galaxy Evolution Explorer (GALEX), a new window
on star formation in galaxies is now available.
The addition
of UV helps to break the age$-$extinction degeneracy in population
synthesis modeling (e.g., \citealp{smith08}).
Furthermore,
since the UV traces 
somewhat older and lower mass stars ($\le$400 Myrs; O to early-B stars) than 
H$\alpha$ ($\le$10 Myrs; early- to mid-O stars), it provides a measure of star formation
over a longer timescale than H$\alpha$ studies.
GALEX imaging of interacting systems 
have shown that tidal features
are sometimes quite bright in the UV (e.g., \citealp{neff05}).
In some cases, 
tidal features previously thought to be purely gaseous 
have been detected by GALEX (e.g., \citealp{hancock07}).
In other systems, GALEX images
have been used to identify new tidal features 
(e.g., \citealp{boselli05}).

To address these issues, we have used the 
GALEX telescope to image a 
well-defined sample of more than three dozen strongly 
interacting galaxies in the ultraviolet.
Combined with broadband optical data,
these images provide information about the star formation
history and dust extinction within the galaxies.
For four of the galaxies in this sample (Arp 82, Arp 284, Arp 285, 
and Arp 305), we have already published
the GALEX images as part of 
detailed studies of the distribution of star formation within
the galaxies, and compared with numerical simulations of the interaction
\citep{hancock07, hancock09, smith08, peterson09}.
The
GALEX images of Arp 24, 85, and 244 were previously analyzed by
\citet{cao07}, \cite{calzetti05}, and \citet{hibbard05}, respectively.
In the current paper, we present the full GALEX dataset for the entire sample, compare with optical data,
discuss global and tidal properties, and provide a brief discussion
of each system.
In a followup paper, we compare with a sample of normal galaxies.

The galaxies in our interacting
sample were selected to be relatively isolated
binary systems, thus they are less complex
than many of the interacting systems studied recently by GALEX,
for example, the Hickson Group studies by \citet{deMello08}
and \citet{torres09}, and the ram-pressure-stripped
NGC 5291 system \citep{boquien07}.
Our galaxies were selected to be relatively simple systems,
thus are more amendable to numerical modeling and detailed
matching of simulations with multi-wavelength datasets.  
In the current paper, we provide a summary of the GALEX data
for the full sample. 

\section{The Interacting Galaxy Sample } 

Our galaxy sample was selected from 
the Arp Atlas of Peculiar Galaxies (Arp 1966),
based on the following criteria:
1) They are relatively isolated binary systems; we 
eliminated merger remnants, close triples, and multiple
systems in which the galaxies have similar
optical brightnesses (systems with additional
smaller angular size companions were not excluded).
2) They are tidally disturbed.
3)
They have
radial velocities less than $<$ 10,350 km/s (most are $<$ 
3500 km/s). 
4) Their total angular size is $^{>}_{\sim}$ 3$'$, to allow for 
good resolution with
GALEX.   
One of our systems, Arp 297, consists of two pairs at
different redshifts.  These two pairs are included separately
in our sample.  
We also include the nearby pair NGC 4567, which fits
the above criteria but
is not in the Arp Atlas.

After removing objects with too-bright stars in their field that
could not be avoided by shifting the target position, our sample consists
of 42 pairs of galaxies.   These systems are listed in Table 1.
Of these pairs, 13 were already 
reserved by GALEX guaranteed time
projects.   We observed the remaining 29 systems (see Section 3).
We then 
combined our new observations with the archival data for 
the other 13 systems that had already been observed.

This GALEX sample overlaps with the sample that we studied in
the infrared using the Spitzer telescope \citep{smith07},
however, it is not identical.   
Spitzer observations were made of some of the galaxies
that we were not able to observe
with GALEX because of 
UV-bright stars in the field.
In addition,
the Spitzer survey omitted galaxies
with angular sizes of the individual disks less than 30$''$, while
these galaxies were 
included in the GALEX study.

\section{Observations}

Table 2 provides the dates, exposure times, and tile names for
the GALEX observations of our sample, including both
our new observations and the archival observations.
When possible, 
we imaged each galaxy for $\ge$1500 seconds in both the far-ultraviolet
(FUV) and 
the near-ultraviolet (NUV) broadband filters of GALEX, which 
have effective bandpasses of 
1350 $-$ 1705\AA~ and
1750 $-$ 2800\AA, 
respectively.
As shown in Table 2, some systems were observed only in NUV or only
in FUV, due to a bright star in the field.
For the archival study, we only selected galaxies with at least
800 seconds total exposure in either the FUV or NUV. 
The GALEX field of view is circular with a diameter of 1.2 degrees.
The pixel size is 1\farcs5, and the spatial resolution is $\sim$5$''$.
In some cases, the GALEX target position was offset from
the position of the galaxy in order to avoid a nearby bright star.

\section{Other Data}

Of our 42 systems with GALEX data, 29 also have broadband
optical images available from the Sloan Digitized Sky Survey (SDSS;
\citealp{abazajian03}).
These galaxies are identified in Table 2.
The SDSS ugriz filters have effective wavelengths of
3560\AA, 4680\AA, 6180\AA, 7500\AA, and 8870\AA, respectively.
Of our GALEX sample of 42 pairs, 
31 have broadband Spitzer 3.6, 4.5, 5.8, 8.0, and 24 $\mu$m
images available.  Most of these have been published
in \citet{smith07}.
Of our 42 systems, 21 have published 21 cm HI maps,
while 23 have H$\alpha$ maps available, either from the literature
or our own unpublished observations with the Southeastern
Association for Research in Astronomy (SARA) telescope.
We also have acquired new optical and Spitzer spectra of a few of the star
forming regions in some of the tidal 
features in our sample.  These will be discussed
in later papers \citep{hancock10, higdon10}.

\section{Overview of Morphologies}

In 
Figures 1 $-$ 14 we show
mosaics of the UV and optical images
for the sample galaxies that have both GALEX and SDSS
images.
These galaxies are displayed in order by Arp number,
and are discussed individually in Section 7.
Figures 15 $-$ 18 show the GALEX images for the systems not observed
by SDSS.
Figures 1 $-$ 18 show that 
the sample galaxies have a large range of collisional morphologies, 
including 
M51-like systems, wide pairs with long tails and/or bridges, 
wide pairs with short tails, close pairs with 
long tails, and close pairs with short tails.
Our sample also includes a possible `Taffy' galaxy, Arp 261,
apparently produced by a near head-on collision
between two equal-mass gas-rich galaxies, as well
as possible previously unidentified ring galaxies in Arp 112, 192,
and 282.

Within the tidal features, we see a range of star formation
morphologies.   In many systems, we see examples of
the so-called `beads on a string' morphology, in which 
regularly-spaced clumps of star formation are seen along
spiral arms and tidal features.
These clumps are generally spaced about 1 kpc apart, which is
the characteristic scale for the gravitational collapse
of molecular clouds \citep{elmegreen96}.
Examples of such `beads' are seen in Arp 34, 35, 65, 72, 82,
84, 86, 100, 242, and 285.

In a few systems, we see very luminous star forming regions
at the base of a tidal feature.  We call these features
`hinge clumps' \citep{hancock09}.   These lie near the
intersection of the spiral density wave in the inner disk
and the material wave in the tail.   These may form
when dense material in the inner disk gets pulled out
into a tail.   This lowers the shear, which may allow
more massive clouds to gravitationally collapse.
Hinge clumps are seen in Arp 65, 72, 82, 242, 270, and 305.

Our sample also includes some candidate `tidal dwarf galaxies' (TDGs),
massive concentrations of
young stars near the tips of tidal features.
The prototypical TDG in the northern tail of Arp 105
\citep{duc97} is included in our sample, along
with the well-studied TDGs in Arp 244 and Arp 245
\citep{mirabel92, duc00}.
In Arp 242, candidate TDGs are seen in both tails.
In Arp 112, in addition to the two main galaxies a third
fainter galaxy is seen, which may be either a TDG, a background
galaxy, a portion of a collisional ring, or
a pre-existing dwarf.
In addition, we have identified possible TDGs in 
Arp 305 \citep{hancock09}, Arp 181, and Arp 202.
Faint UV clumps are also visible near the end of the long HI
tail south of Arp 270, but no optical redshifts are available at 
present to confirm that these are associated with the tail.
These candidate TDGs are discussed in detail in Section 7.

Our sample also contains numerous examples of accretion
from one galaxy to another.   One of the best-studied
examples is the northern tail of Arp 285,
which was likely produced from material accreted from the 
southern galaxy
\citep{toomre72, smith08}.
According to our numerical simulations, the material in this
tail fell into
the gravitational potential of the northern galaxy, overshot
that potential, and is now gravitationally collapsing and forming
stars \citep{smith08}.
We call such features `accretion tails', to distinguish
them from classical tidal features.  The inner western
tail of Arp 284 was likely produced by the same 
process \citep{struck03}.   The southern tail
of Arp 105 may have formed by
the same mechanism, as well as the northwestern tail
of Arp 34.  Arp 269 may be an example of accretion from one
galaxy to another; alternatively, it may be an example
of ram pressure occurring during the passage of one galaxy
through the gaseous disk of another.
Arp 87 contains a good example of a polar ring-like system,
caused by accretion from a companion.
These systems are described in more detail in Section 7.

Our sample contains only a few tidal features that 
have high HI column densities ($\sim$4 $\times$ 10$^{20}$ cm$^{-2}$)
but are not detected in
our GALEX maps or published optical maps:
Arp 84, 
Arp 269, Arp 270, and Arp 280.
A few additional tidal features have somewhat lower HI column
densities, between 6 $\times$ 10$^{19}$ cm$^{-2}$
and 10$^{20}$ cm$^{-2}$, but no GALEX/SDSS counterparts:
Arp 85, 86, and 271.   
Deeper GALEX and optical images are needed to check whether
these are truly starless structures.
These features are discussed in more
detail in Section 7.

\section{UV $-$ Optical Colors}

\subsection{Photometry}

Magnitudes in the various GALEX and SDSS
bands for the main disks and the tidal features
are given in Table 3.
For the systems with possible TDGs near the tips of
tidal features,
we determined the colors of the TDG 
separately from the connecting tail.
These are labeled 
`TDG' in Table 3, although we emphasize that in most cases
it is unclear whether these are truly TDGs.
These features are discussed in detail in Section 7.

To determine these magnitudes, 
we used a set of rectangular boxes that 
covered the observed extent of the 
targeted area in the GALEX and SDSS images, 
but avoided
very bright stars.
These boxes generally coincide with those used in
our Spitzer study \citep{smith07}, but for some 
galaxies these were modified, for example,
if the observed extent of the tidal features or the disk was larger
in the GALEX images.
For each system, the sky was measured in a set of rectangular areas 
without bright stars or galaxies. 
The uncertainties in Table 3 include both the 
statistical uncertainty
and an uncertainty due to variations from
sky region to region, as in \citet{smith07}.
The magnitudes in Table 3 were corrected for Galactic
extinction as in \citet{schlegel98}, using the \citet{fukugita04}
extinction law in the SDSS bands and the \citet{cardelli89} law in
the UV.  The fluxes were converted to magnitudes on the AB system
\citep{oke90} using zero point fluxes of 3631 Jy in each band.

For the three galaxy pairs that span two SDSS 
fields (Arp 85, 101, and 285),
the optical images shown are the mosaicked
SDSS images from \citet{hogg07}.  However, the magnitudes
quoted in Table 3 for Arp 101 and 285 were measured on the
original unmosaicked images.  
For Arp 270 and 297, 
we also had to use adjacent SDSS images
to do the SDSS photometry of the ends of the tails.
For Arp 85, 
the 
angular size is so large that sky measurements on the SDSS images
is problematic, thus we do not provide SDSS magnitudes.

\subsection{Disk vs.\ Tidal Colors, Compared to Population Synthesis Models}

In Figures 19 $-$ 23, we plot various
UV $-$ optical color-color plots for the main bodies and the
tidal features of the Arp galaxies.
The candidate TDGs are plotted separately from their parent tidal
features.  Selected features are labeled in these figures.
On these plots, we superimpose evolutionary
tracks from version 5.1 of the 
Starburst99 population synthesis code \citep{leitherer99}, assuming
instantaneous
bursts,  
\citet{kroupa02} 
initial mass functions, and an initial mass range of 0.1 $-$ 100 
M$_{\sun}$.
This version of the code includes the Padova asymptotic giant-branch
stellar models \citep{vaz05}.
We display both solar metallicity and 0.2 solar metallicity
models
on these plots.
We convolved the model spectra with the SDSS and GALEX
filter response functions
to obtain the model colors shown in Figures 19 $-$ 23.

As in our Arp 285 paper \citep{smith08},
to these model colors
we added in the 
H$\alpha$ line,
which can contribute significantly to the r band flux for very young
ages.  
We note that this effect is redshift-dependent.
For the models plotted
in Figures 19 $-$ 23, 
for the H$\alpha$ line 
we assumed the velocity
of Arp 285, which is typical of the sample
as a whole (Table 1).  At this redshift,
for very young star forming regions
(1 $-$ 5 Myrs), the r magnitude decreases by 1.1 $-$ 0.25 magnitudes
due to the presence of H$\alpha$ \citep{smith08}.
However, 
for our highest redshift objects,
the H$\alpha$ line is shifted to the edge of the r band filter
where the sensitivity is down by a factor of $\sim$2.5.  
For these high redshift galaxies, 
the model r magnitudes plotted
in Figures 19 $-$ 23 for very young ages are too bright.
The effect of H$\alpha$ is illustrated in Figure 21, the 
g $-$ r 
vs.\ 
u $-$ g
plot, in which we show tracks with and without H$\alpha$.
For the rest of the plots, we only display the models which include
H$\alpha$.
Note that the plotted models do not include the [O~III] $\lambda$5007 or
H$\beta$ line, which can contribute substantially to the g filter
for low metallicity young galaxies (e.g., \citealp{kruger95, west09}).

In Figures 24 $-$ 28, we provide histograms of various colors
for the main bodies of the galaxies, the tidal features,
and the candidate TDGs separately.
Selected features are labeled in these figures.
For the main bodies of the interacting galaxies, there is a range
of colors, reflecting a range in star formation histories, 
extinction, and progenitor morphological types.
For example, the four reddest disks in FUV $-$ NUV are
all apparently early-type galaxies:
Arp 173 N (classified as S0 in NED\footnote{The NASA Extragalactic Database;
http://nedwww.ipac.caltech.edu}),
Arp 100 S = IC 19 (classified as E in NED), 
Arp 290 S = IC 195 (classified
as SAB0 in NED), 
and Arp 120 N = NGC 4435 (classified as SB0 in NED).
The UV colors of the fifth reddest disk in FUV $-$ NUV,
Arp 89 W 
(= KPG 168 B), may be strongly affected by
extinction, since it 
is an edge-on disk galaxy classified as Sc in NED.
For comparison,
in the Nearby Galaxies Atlas \citep{gildepaz07}, there is
a correlation of FUV $-$ NUV with morphological type, with
the early-type galaxies (E/S0) being
redder, with FUV - NUV between 1 $-$ 2,
consistent with our reddest systems.

In contrast, many of the bluest disks 
in our interacting sample
have very late morphological
types according to NED.  For example,
in NUV $-$ g the bluest disks are 
Arp 202 S (Im pec), Arp 24 main (SABm), and 
Arp 305 N (SBdm).  These may have originally been late type
galaxies, or, 
alternatively, their morphology
may have changed due to the encounter.

We do not find a large difference in the 
colors of the tidal features 
compared to those of the disks, on average (see Figures 24 $-$ 28).
These results suggest that, on average, the tidal features do not
have stronger mass-normalized
bursts of star formation than the disks. 
Kolmogorov-Smirnov (K-S) tests cannot rule out the possibility
that the colors of the disks, the tails/bridges, and the TDGs
come from the same parent population.  
The most significant difference is found
for g $-$ r, where the tidal features are slightly
bluer than the disks, and a K-S test 
gives a 2.5$\%$ probability that the two
samples come from the same population. 
This result is suggestive, but inconclusive.
\citet{schombert90}
also found possibly bluer B $-$ V colors (approximately g $-$ r) 
for the tidal features
than the disks, while their V $-$ i colors (approximately r $-$ i)
for the main bodies and tidal features
were similar. 
We note that the \citet{schombert90} sample contains more early-type galaxies
than our sample.
It also contains a number of merger remnants,
which were excluded from our sample.   The main
bodies of these galaxies are likely redder 
on average
than our sample galaxies 
due to older stars or 
more extinction.

Comparison of the colors of the tidal features with the
population synthesis models (Figures 19 $-$ 23) show that, in general,
the light from both the disks and
the tidal features is not dominated by
very young stars, and these features are not completely
extinction-free.   As with the main bodies of these 
galaxies, there is a range of colors for the tidal features,
due to a range in star formation properties and original
morphologies.  For example, the three tails that are reddest in
FUV $-$ NUV are the two broad diffuse tails of Arp 283
and the southern tail in Arp 173, which
extends from a red S0 galaxy.
No clumps of star formation
are visible in these tails.
In contrast to these features, 
the bluest tail in FUV $-$ NUV is Arp 120 WT,
with FUV $-$ NUV = $-$0.03.   
This feature was originally discovered
by \citet{boselli05} using these same GALEX
images.  

\subsection{The Colors of the Tidal Features vs.\ their Parent Disks }

Another useful test is to compare the colors of the
individual tidal features with the colors of
their own parent disk, rather than with the sample as
a whole.
In Figure 29, in the various colors
we provide histograms of the differences 
between the 
colors of the disk and 
the colors of their matching tidal features.   For the bridges,
we matched with the most probable progenitor disk of the two
galaxies.  
We matched 
the northern Arp 285
tail and the southern Arp 105 tail
with 
the southern and northern galaxies in those pairs, respectively,
as those are the most likely progenitors (see Section 7).

For most of the histograms in Figure 29, the distributions
of color differences 
peak near zero, 
and the mean color difference is smaller than the average
measurement uncertainty.
For NUV $-$ g, however, there is a significant shift
of the histogram to the right of zero,
with the mean (NUV $-$ g)$_{disk}$ $-$ (NUV $-$ g)$_{tidal}$
color being 0.26, compared to a typical measurement uncertainty in this
color of 0.03 magnitudes in Table 3.   Thus the tidal features are, on average,
bluer in NUV $-$ g than their own parent disks.
This confirms that, on average, tidal
features are indeed more prominent in the UV than in the optical,
compared to their own parent disks.
For g $-$ r, there is a marginal effect, in that 
the mean (g $-$ r)$_{disk}$ $-$ (g $-$ r)$_{tidal}$ is 0.09,
while the average g $-$ r measurement uncertainty per feature is 0.03.
We note that the Figure 29 colors for the tidal features do not
include the fluxes of the candidate tidal dwarf galaxies, which were
measured separately.  As seen in Figure 25, for Arp 181, 202, and 305,
the TDGs are considerably bluer in NUV $-$ g than their parent tidal
features.   The inclusion of these regions as part of the tidal feature
would increase the difference in color from their associated disk.

Inspection of the population synthesis models
plotted in Figure 20 shows that a shift to the blue
for NUV $-$ g without
a strong change in g $-$ r is likely an age effect rather than
an extinction effect.   With the GALEX data, we are able
to break the age-extinction degeneracy to a certain extent.
Thus we suggest that the stars in
the tidal features are younger, on average, than those 
in their parent galaxies, and star formation, rather
than dust extinction, is the primary factor responsible for the bluer
NUV $-$ g colors.   

In addition to age,
another factor that may be important is
metallicity.   
In general, lowering the metallicity makes the 
NUV $-$ g colors bluer (see Figure 20).  Of all the colors
investigated in this study, NUV $-$ g is the most strongly affected
by metallicity.
Since tidal features tend to
be drawn from lower metallicity regions in the outer disks,
tails and bridges likely have lower abundances than their
parent galaxies on average.   Thus the observed
difference may be due in part to lower metallicities in the tidal
features.   
Only a handful 
of the tidal features in our sample have available oxygen abundances, which
range from log(O/H) + 12 of
8.4 to 8.7, or
$\sim$1/3 $-$ 1/2 solar (see Section 7).
More direct measurements of abundances in
our sample galaxies 
would be useful
to better distinguish between
age and metallicity as the cause
for this color difference.

\subsection{Colors of Subsets of the Sample }

As noted in Section 5, our sample is comprised of pre-merger
interacting pairs of galaxies with a large variety of morphologies.
The sample includes both close pairs and wide pairs, equal mass
pairs as well as unequal mass pairs, and pairs with long tails
as well as pairs with short or weak tails.  To investigate whether 
any of these subtypes of interacting galaxies stands out from
the rest of the sample in their UV/optical colors, we have 
selected four different subsets from the sample, two selected based 
on morphology, and two based on pair separation.   First, we
divided the sample into two groups based on pair separation 
with the `wide pairs' being the 15 pairs with separations greater than
or equal to 30 kpc, and the `close pairs' sample being the 27 galaxies
with separations less than 30 kpc (see Table 1).
Next, we created a subset including the 10 M51-like 
galaxies in the sample (see Table 1).  Finally, we separated out
another subset that includes the 10 pairs with disturbed disks
but with weak or no tidal features visible in the optical/UV images.
In Figure 30, we present
histograms of FUV $-$ NUV,
NUV $-$ g, and g $-$ r colors
of the galactic disks for these subsets of galaxies.
We used K-S tests to search for significant differences in the 
distributions of optical/UV
colors of 
these subsets of galaxies.  In 
no case were we able to rule out the hypothesis that they came from
the same parent population.   

These results are consistent with our earlier study on the Spitzer
colors of these galaxies,
in which
we could not find significant differences 
between wide and close pairs, and
between M51-like systems and the rest of the sample
\citep{smith07}.
In contrast to our results for the SB\&T galaxies, 
earlier studies based on H$\alpha$
equivalent widths and optical spectroscopy
detected a significant difference in star formation
rate between close and wide pairs, though with a large amount
of scatter \citep{barton00, lambas03, nikolic04}.  
Our different conclusions are likely caused in part to selection effects.
Our pairs were selected based on the presence of strong tidal distortion.
Many of our widely-separated
pairs have very long tidal features, which indicates that they 
probably were closer together at some point in the past,
and have already undergone a strong gravitational encounter.
In contrast, in the earlier
studies, pairs were selected based solely on proximity in space,
and thus many of the wider pairs may not have experienced such strong tidal
disturbances.  Another factor is simply sample size, as our subsets
of galaxies are quite small, compared to these earlier studies.
Larger 
sample sizes are needed to further
investigate whether particular classes of interacting galaxies
are more likely to have enhanced UV/optical colors than other types.

\subsection{UV/Optical Colors vs.\ Spitzer IR Colors}

Broadband colors in both the UV/optical range and in the mid-infrared 
regime
are sometimes used as indicators of recent star formation in galaxies,
however, each method has advantages and disadvantages. 
Younger stars have bluer UV/optical colors, however, dust extinction
and/or metallicity differences can redden these colors, confusing the issue.
The Spitzer 3.6 $\mu$m broadband filter is generally assumed
to be dominated by the stellar continuum from the older
stellar population, while the 24 $\mu$m
band is dominated by emission from hot dust heated by UV
photons from young stars, thus redder [3.6 $\mu$m] $-$ [24 $\mu$m] color
are associated with
younger stellar populations on average (e.g., \citealp{smith07}).  
This is also the
case for the [3.6 $\mu$m] $-$ [8.0 $\mu$m] color, however, in
addition to hot dust the Spitzer
8 $\mu$m band also includes the prominent
interstellar polycyclic aromatic hydrocarbon (PAH) features, which 
can vary from galaxy to galaxy
depending upon the chemistry and the hardness of the UV radiation field.
In addition, in extreme starbursts, interstellar
contributions can be significant in the 3.6 $\mu$m Spitzer band \citep{smith_hancock09}.

To determine how well 
the colors in these two wavelength regimes 
correlate, in Figure 31 we compare the Spitzer [3.6] $-$ [24] and [3.6] $-$ 
[8.0] colors
for the disks and tidal features in our sample with
their GALEX/SDSS FUV $-$ NUV, NUV $-$ g, and g $-$ r colors.  
The Spitzer magnitudes were obtained from \citet{smith07} when possible.  
For the
galaxies in our current sample that are not in \citet{smith07}, when Spitzer
images were available in the Spitzer archives, we downloaded
these images and extracted magnitudes for the appropriate regions. 
As noted earlier, in most cases, the GALEX/SDSS magnitudes
were measured over the same area of the sky as the \citet{smith07}
Spitzer magnitudes.
However, 
in the current study we measured the GALEX/SDSS magnitudes
of candidate TDGs separately from their host tidal feature; this was not
done for the Spitzer fluxes in \citet{smith07}.  
Also,
in a few cases we modified the Spitzer regions somewhat.
For these galaxies, we re-determined
the Spitzer fluxes in regions that match the regions used for the 
GALEX/SDSS fluxes.

Very weak trends are apparent in some of the
panels of Figure 31, in that galaxies that are bluer in
NUV $-$ g and g $-$ r tend to have redder
Spitzer [3.6] $-$ [24] and [3.6] $-$ [8.0] colors.   
However, a very large amount of scatter is present in these plots,
and there are galaxies with both red Spitzer colors and red UV/optical colors.

There are several likely reasons for the large scatter in these plots.  
One factor
is system-to-system variations in the geometry of the stars and dust, 
leading to larger UV/optical extinctions in some systems than others, 
for the same
dust emission.  In addition, 
the dust properties, the metallicities, and the PAHs vary from galaxy to galaxy,
along with the strength and hardness
of the interstellar radiation field.
Furthermore, these fluxes were measured over large regions within the galaxies,
entire disks or tidal features,
thus they are the sums of the fluxes from multiple star forming regions with different 
ages and extinctions, combined with the emission from more quiescent regions.

This latter point is illustrated in Figure 32, in which we plot
[3.6] $-$ [8.0] vs.\ NUV $-$ g and g $-$ r for individual star forming
clumps within Arp 285, using the photometry from \citet{smith08}.
The magenta open diamonds are points from the northern tail, the 
cyan open circles are clumps in the NGC 2856 disk, and 
the black open squares are clumps in the NGC 2854 disk.
These clumps show essentially constant, and very red, [3.6] $-$ [8.0] colors
for the clumps, implying very young stellar populations.
In contrast, there is a large spread in NUV $-$ g and g $-$ r for these
clumps, indicating a large range of extinction in these clumps.  
From population synthesis of the UV/optical colors for the individual
clumps, we determined extinctions that range from E(B $-$ V) = 0.1 to 0.3
for the clumps in the tidal features, to 0.2 to 1.3 for the disk clumps
\citep{smith08}.
For most of these clumps
the inferred stellar ages are quite young, 4 $-$ 20 Myrs, but
with large uncertainties.

Comparison of Figure 32 with the two upper right panels in Figure 31
show that individual star forming regions populate a different part
of these color-color plots than galaxies as a whole.  This is
likely due to significant contributions from the underlying
older stellar population to the global Spitzer 3.6 $\mu$m flux densities,
as well as to the broadband optical fluxes.   This example demonstrates
the importance of spatially-resolved measurements in interpreting
UV/optical/IR colors in terms of extinction and stellar ages in galaxies.

\section{Notes on Individual Systems}

\noindent{\bf Arp 24:} Arp 24 (Figure 1) is an M51-like system, but
with only a weak bridge.   In spite of the very long 
exposure with GALEX of 85,000 sec, no very extended UV emission
is observed.
No HI map is available for this system.
A detailed analysis
of the star forming regions in Arp 24, 
utilizing the GALEX images
as well as the Spitzer maps and optical data,
was published by \citet{cao07}.

\noindent{\bf Arp 34:} 
Arp 34 (Figure 1) contains three galaxies, a
pair NGC 4614/5 with approximately equal optical luminosity,
and a third fainter galaxy NGC 4613
to the northwest.
The most southern of these galaxies, NGC 4614, has a prominent
outer ring in the GALEX and optical images.
The brightest of the galaxies,
NGC 4615, has prominent `beads' of star formation in
its inner disk in the GALEX images. It also has 
an unusual extension to the northwest, which extends
out from the disk at an angle.
This feature 
is detected in the H$\alpha$ map of \citet{gavazzi03, gavazzi06},
thus it is at the same redshift as Arp 34.  The
tip of the tail is quite bright in H$\alpha$
and is faintly visible in the 
Spitzer 8 $\mu$m image (Smith et al.\ 2007).
This feature is 
also quite blue
in NUV $-$ g.   
The similarity of this feature to the Arp 285 accretion tail
\citep{smith08} 
suggests that it may also have been caused by accretion,
however,
unlike Arp 285, 
there is no obvious optical bridge between the two galaxies,
and 
no HI map is available at present for Arp 34.
NGC 4615 also has a second tail extending to the south
towards NGC 4614, which is quite 
blue in
NUV $-$ g.  

\noindent{\bf Arp 35:}
Arp 35 (Figure 15) is 
a widely separated M51-like system. The brighter northern
galaxy has strong tails
with bright knots of star formation, with a hinge clump at
the base of the southern tail.
The southern galaxy has two
short tails.
The southern tail of the smaller southern galaxy
is the second bluest tidal feature in FUV $-$ NUV, after
the northern tail of Arp 120.   Arp 35 was not
observed by SDSS or Spitzer, thus little additional
information about this tail is available.  It is possible
that the UV clump in this tail is an `accretion knot', as
in Arp 285.  Alternatively, it may be a background quasar. 
Obtaining an optical spectrum of this source
would be very valuable in determining its nature.

\noindent{\bf Arp 65:}
Arp 65 (Figure 15) is a widely separated equal mass pair of spiral galaxies.
The long northern tail of the western galaxy has a prominent 
hinge clump 
near the base of the tail, which shows up brightly
in the 8 $\mu$m Spitzer map \citep{smith07}.
Beads of star formation
are also visible further along
this tail.
In the southern tail of this galaxy, 
an offset between the old and young stars is seen in
the 
Spitzer images \citep{smith07}.

\noindent{\bf Arp 72:} 
The GALEX images of 
Arp 72 (Figure 2) show a long `beaded' tail extending to
the east of the main galaxy.
This tail is also seen in 
the SDSS and Arp Atlas optical
images, but is more pronounced in the GALEX maps and 
has 
very blue
NUV $-$ g colors.
The main galaxy of Arp 72, NGC 5996, 
shows a bar-like structure in the SDSS
images, connecting to a prominent spiral arm 
which extends to the
north.  
The bridge that connects NGC 5996 to the 
low mass
companion NGC 5994 is very bright in the UV, with
a prominent hinge clump near NGC 5596.
The central portion of this bridge has an oxygen abundance
of log(O/H) + 12 $\sim$ 8.7 \citep{hancock10}.
In our SARA H$\alpha$ map, the 
bridge and the
northern spiral arm are quite bright, but the eastern
tail was not detected.

\noindent{\bf Arp 82:}  
We have already published the 
GALEX images of the M51-like galaxy pair Arp 82,
along with H$\alpha$ and Spitzer
images
\citep{hancock07}.
The GALEX maps (Figure 2) show a long clumpy tail to the north and
a prominent bridge between the two galaxies, as well
as a faint arc of UV emission to the east.   This 
arc had previously been seen in 21 cm HI maps by 
\citet{kaufman97}.   Star forming regions within
this arc are also visible in the SDSS maps.
The colors of this arc are not unusually blue 
in the SDSS/GALEX maps, compared to other tidal
features in our sample.

\noindent {\bf Arp 84 (`The Heron'):} 
Arp 84 (Figure 3) consists of a pair of unequal mass spiral galaxies
connected by a bridge formed from material
pulled out of the smaller northern galaxy
NGC 5394.
NGC 5394 also has a second tail extending to
the north. Along the inner portion of
this tail, a series of `beads' is visible
in the SDSS image.
The most northern knot along this `string' is bright in
the GALEX image, as well as in the Spitzer 8 $\mu$m map
\citep{smith07} and in H$\alpha$ \citep{kaufman99}.
It is unclear whether this northern
tail is solely a classical tidal tail,
or if there is a second
`accretion component' 
interacting with the tidal gas.  Perhaps there
are two colliding components in this tail which
are triggering star formation, as with the Arp 284
bridge \citep{struck03}.

The larger southern
galaxy in Arp 84, NGC 5395,
also has two tails.   
The northern tail of the southern
galaxy 
is visible in GALEX as well as in the \citet{kaufman99} HI map.
This feature is
called the `arm extension' by \citet{kaufman99}.
A bright star in the vicinity makes the GALEX and SDSS photometry
of this feature uncertain.

The southern tail of NGC 5395 is visible in the 
\citet{kaufman99} HI map of the system as a large gaseous
plume extending to the 
south of the galaxy.
This HI feature
is not clearly detected in the GALEX maps, in spite of
its relatively high HI column density of 
N$_{HI}$ $\sim$ 4 $\times$ 10$^{20}$ cm$^{-2}$, 
and relatively narrow HI line width of
$\sim$100 km~s$^{-1}$ \citep{kaufman99}.  In the GALEX map,
no diffuse UV emission is seen in this HI plume,
although
a few UV-bright clumps are present.
No redshifts are available at present
for these clumps,
thus it is uncertain whether they are associated with the
galaxy, or are foreground/background objects.
The numerical simulation shown
in \citet{kaufman99} does not produce this plume, which suggests
that it may have formed during an earlier encounter.

\noindent{\bf Arp 85 (M51; `The Whirlpool'):} 
GALEX images of Arp 85 (Figure 3) were previously analyzed 
by \citet{calzetti05}, in combination with Spitzer infrared
and Hubble Space Telescope optical images.
Longer exposure GALEX images are now available (see Table 2), but 
the 
long gaseous
tail visible to the south of Arp 85 \citep{rots90} is still undetected.
This is not surprising, as this feature 
has a relatively
low HI column density of 6 $\times$ 10$^{19}$ cm$^{-2}$,
thus may not have a stellar component.
Diffuse UV emission is seen to the north of NGC 5195, the smaller 
galaxy in the pair.
In the FUV map, a row of star forming regions is
visible in the bridge and into NGC 5195, however, the
underlying disk of NGC 5195 is not visible.

\noindent{\bf Arp 86:} 
The GALEX images of the M51-like pair Arp 86 (Figure 15) show a 
clumpy arc of UV emission to the south of the main galaxy, connecting
to the companion.  
This feature is similar to the eastern arc of 
Arp 82 \citep{kaufman97, hancock07}.
This arc lies within the HI disk in
the HI map of Arp 86 published by \citet{sengupta09}.
This loop is
also faintly visible in the \citet{arp66} Atlas
image (this galaxy was not observed by the SDSS).

The HI map of Arp 86 shows a strong tail to the 
north, looping to the west \citep{sengupta09}.
The base of this tail is seen in the GALEX NUV image.
Near the tip of this tail to the northwest, a concentration
of gas with N$_{HI}$ $\sim$ 10$^{20}$ cm$^{-2}$ is 
visible in the HI maps.   This is not detected in the 
GALEX images.

The compact dwarf galaxy 2MASSXJ23470758+2926531
is visible in the GALEX images
to the southeast of the Arp 86 system.
There is an HI counterpart to
this galaxy, showing that it is at the same redshift as Arp 86 
\citep{sengupta09}.
The HI maps also show a concentration of gas to the south of the companion.
A UV source is visible in the vicinity of this HI cloud 
in the GALEX images, but whether it is at the same redshift
as Arp 86 is unknown.

\noindent{\bf Arp 87:} 
In the SDSS images of the equal-mass disk galaxy pair Arp 87 (Figure 4),
a swirl of blue stars 
in a polar ring-like structure
encircles the edge-on disk of the northern galaxy.
These stars likely formed from gas accreted from the southern galaxy
along the bridge.   Another arc of young stars is visible to the
north
of the southern galaxy, probably formed from gas that was pulled
out into the bridge and is now falling back on the galaxy.

Two additional small disk galaxies are visible in the Arp 87 field,
northeast of the northern galaxy and southeast of the southern galaxy.
No redshifts are available for these objects, thus it is unknown whether
they are associated with Arp 87.

\noindent{\bf Arp 100:}
Arp 100 (Figure 15) is a widely separated elliptical/disk galaxy pair.  The northern
galaxy has two long tails, with four prominent `beads' visible
along the brighter northern tail
in the GALEX images.

\noindent{\bf Arp 101:} 
Arp 101 (Figure 5) is another wide elliptical/disk galaxy pair,
with a long tail extending to the north of
the northern disk galaxy.
Between the two galaxies, a very broad diffuse bridge
is visible in the SDSS images.
This is also faintly visible in the \citet{arp66} Atlas picture,
but not in the GALEX images.

\noindent{\bf Arp 105 (`The Guitar'):} 
Arp 105 contains one of the best-studied examples
of a possible TDG, which lies at the end of a long tail
extending to the north of a disturbed spiral.
This source is named the `TDG' in Table 3.
The spiral in turn is connected 
by a bridge
to an elliptical 
galaxy to the south of the spiral.   
Further south of the elliptical 
is a bright knot of star formation 
\citep{stockton72} which has also been considered a possible
TDG \citep{duc97}.
In HI maps, both the northern TDG candidate and the southern
knot of star formation are luminous, containing 7 $\times$ 10$^9$ M$_{\sun}$
and 5 $\times$ 10$^8$ M$_{\sun}$ of HI, respectively
\citep{duc97}.
Kinematic studies suggest that the northern TDG in Arp 105 
may be due in part to a projection effect, while 
the southern star forming region may be a rotating self-gravitating
object
\citep{bournaud04}.
The northern TDG candidate has log(O/H) + 12 = 8.6, while the
southern knot of star formation has log(O/H) + 12 = 8.4 \citep{duc94}.

In the GALEX images (Figure 5), 
the spiral and the northern TDG are quite bright in the UV,
but the highest UV surface brightness is found in
the knot of star formation south of the elliptical.
We suggest, based on analogy to Arp 285 \citep{smith08} and 
proximity to the elliptical, that the southern
star forming region in Arp 105 is an accretion tail, 
rather than
simply a classical tidal tail coincidently seen in projection
behind the elliptical.
We note that in H$\alpha$ maps \citep{bournaud04} this source
resolves into three `beads' of star formation.
In Table 3, this feature is simply called the `southern tail'.

Another galaxy
northeast of the Arp 105 spiral,
VV237d, 
also shows up brightly in the GALEX images.
This galaxy has a velocity that differs
from that of the Arp 105 spiral
by 2700 km~s$^{-1}$, 
thus is probably
a foreground galaxy.

\noindent{\bf Arp 107 (`The Smile'):} 
In the SDSS image (Figure 6),
the main disk of 
Arp 107 shows a prominent spiral arm/partial 
ring-like structure, connected to a short tail.
This galaxy is connected to
the elliptical-like companion 
by a bridge.   
The basic morphology of this system can be produced
by a hybrid model, in between a classical ring galaxy
simulation of a head-on small impact parameter collision
and a prograde planar
encounter \citep{smith05a}.
In the Spitzer 8 $\mu$m map,
the ring is quite bright, especially in the south, 
as are two luminous knots of star formation in
the northwestern sector of the system \citep{smith05a}.
The ring and the northwestern
knots of star formation are also very bright in
the GALEX images, while
the companion is relatively faint.  
In contrast,
the northern plume of the northern galaxy has a red NUV $-$ g color (Table 3)
and a blue Spitzer [3.6] $-$ [8.0] color \citep{smith07}, thus
it contains mostly older stars.
There is a bright foreground star in front of
the Arp 107 main disk, between the ring and 
the nucleus \citep{smith05a}.  This is clearly visible
in the SDSS color image as a red point source.

\noindent{\bf Arp 112:} 
In the SDSS image of Arp 112 (Figure 6), three 
galaxies are visible: a large angular size disturbed spiral
NGC 7806, a more compact
galaxy to the southwest named 
NGC 7805, 
and a fainter
arc-like structure to the east, named 
KUG 2359-311 (called `TDG' in Table 3).
In the Arp Atlas picture, KUG 2359-311 
appears to be
attached to NGC 7806 by a faint tail in
the south.  There also may be
a faint connection between these two galaxies in the north.  However,
these features
are not visible in the SDSS and GALEX images.
No
redshift is available yet for 
KUG 2359-311,
thus it is uncertain
whether it is part of the Arp 112 system or not.  
KUG 2359-311
is not particularly blue in the UV$-$optical colors
(see Figure 24 $-$ 28).
If KUG 2359-311 is at the same redshift
as the rest of Arp 112,
it may be either a pre-existing dwarf galaxy or
a tidal dwarf galaxy.
Alternatively, 
NGC 7806 and KUG 2359-311 may be the remains of
a ring galaxy produced by a collision with
NGC 7805.
Deeper optical and UV images as well
as optical spectroscopy are needed to 
better determine the nature of
these galaxies.

In addition to the possible connection to 
KUG 2359-311,
NGC 7806 has a strong tail extending to the north with
two UV-bright clumps.  Redshifts are also needed for these
sources, to determine whether these are associated with the system.

\noindent{\bf Arp 120:}
The Arp Atlas and SDSS images
of Arp 120 (Figure 7) show a 
disturbed spiral
NGC 4438
and an E/S0
galaxy NGC 4435.
Tail-like features are visible 
in the Arp photograph extending to 
the west and north of NGC 4438.
NGC 4438 and NGC 4435
were originally thought to have undergone
a near-head-on collision \citep{kenney95}.
However, subsequent deep H$\alpha$ imaging showed
filaments of ionized gas connecting
NGC 4438 and the giant elliptical M86 to
the west, suggesting that instead
there was an NGC 4438/M86 encounter
\citep{kenney08}.
The GALEX images included in the current study
have already been published by
\citet{boselli05}, who identified a new tidal
feature to the northwest of NGC 4438. 
This feature (called `WT' in Table 3) is 
the bluest tail in FUV $-$ NUV in our sample,
with FUV $-$ NUV = $-$0.03.   

\noindent{\bf Arp 173:}
A long tail is visible to the south of
the southern galaxy in Arp 173 (Figure 7).   This tail has
very red UV/optical colors, perhaps in part because
its progenitor is an S0 galaxy with little star formation.

\noindent{\bf Arp 181:} 
Arp 181 (Figure 8) contains two approximately equal mass spiral galaxies,
NGC 3212/5.  The galaxy to the northwest, NGC 3212, 
has a long tail extending to the west.
A clump is visible 
near the end of this tail in the \citet{arp66}
and SDSS optical images, as well as the GALEX
images.   This clump (called `TDG' in Table 3)
has very blue optical/UV colors (NUV $-$ g = 1.1; one of the
bluest in our sample),
however, no optical spectrum
is available and it was not detected in our SARA H$\alpha$ map,
thus it is unclear whether it is at the same redshift as
Arp 181.

Further west,
beyond the end of the tail
at the edge of the Arp Atlas photograph,
another galaxy is visible (called the `far west' galaxy in Table 3).
This does not have any obvious link to the
tail in any of the images. 
We have an optical spectrum of this galaxy \citep{hancock10}, 
which shows that it 
is at the same redshift as Arp 181.  It is marginally detected
in our H$\alpha$ map, and is visible in the Spitzer maps.
In the SDSS image it
looks like a spiral galaxy or a disturbed disk with short tidal tails.  
It is extremely blue in NUV $-$ g (1.2 magnitudes).
Our optical spectrum implies a relatively low oxygen abundance
of log(O/H) + 12 = 7.8.   
This may be either a pre-existing dwarf
galaxy or a recently detached TDG.

\noindent{\bf Arp 192:} 
The peculiar system Arp 192 (Figure 8) has a very long distorted tail
that curves to the south and then to the east.   In the SDSS
images, the main body
of this system appears to be a very close pair, with 
a disk galaxy containing a ring galaxy-like
loop and a second compact galaxy.
In the \citet{arp66} Atlas photograph of Arp 192, a straight
jet-like feature is seen extending to the northwest out of
the second compact nucleus.
This feature is not seen in the SDSS or GALEX images, or in the near-infrared
images of \citet{bushouse92}, thus the Arp Atlas
feature may be an artifact.

\noindent{\bf Arp 202:} 
Arp 202 (Figure 9) is a very close interaction
between an edge-on disk galaxy in the north and a smaller
irregularly-shaped galaxy to the south.
The southern galaxy has quite blue optical/UV colors, and
a long clumpy tail extending
to the west.   The tip of this tail is very prominent in
the GALEX images.
We have an optical spectrum of this source that confirms
it is at the same redshift as Arp 202 \citep{hancock10}, 
thus it is another possible
TDG.
This candidate TDG has very blue UV-optical
colors, suggesting a relatively young age with little extinction.
However,
it was not detected
in our Spitzer 8 $\mu$m map \citep{smith07}
or in our SARA H$\alpha$ map.
This suggests that this feature is in a post-starburst stage, and is not
currently forming stars.  
Our optical spectrum gives an approximately
solar oxygen abundance for this candidate TDG of log(O/H) + 12 = 8.9.

\noindent{\bf Arp 242 (`The Mice'):}
This equal-mass pair with two strong
tails is very well-studied at many wavelengths (e.g., 
\citealp{hibbard96, braine01}).   In the GALEX data (Figure 9), 
three UV-bright
`beads' are visible along the northern tail.  Each is associated
with a peak in the HI map of \citet{hibbard96}.   
The most northern HI concentration has been classified a possible
TDG
based on kinematic information
\citep{bournaud04}.
At the tip of the southern tail, there is another UV-bright
concentration in the GALEX maps, which also has an HI counterpart.
We classify this as a second possible TDG.
Near the base of the southern tail is a UV-bright concentration, another
possible `hinge clump'.

\noindent{\bf Arp 244 (`The Antennae'):} 
The interacting pair Arp 244 (Figure 16) has been studied in
detail by many authors (e.g., 
\citealp{hibbard01, bastian06}).
A concentration of HI and young stars near
the tip of the southern tail of Arp 244 
\citep{schweizer78, vanderhulst79, mirabel92} 
has been considered a possible TDG.
The GALEX UV images of Arp 244 used in the current study
have been previously published by \citet{hibbard05},
and show that both tails are quite bright in the UV.
The southern tail has two
parallel ridges of HI and UV emission that meet at the
TDG \citep{hibbard01, hibbard05}.
This TDG has an oxygen abundance log(O/H) + 12 $\sim$ 8.4
\citep{mirabel92}.

\noindent{\bf Arp 245:} 
The possible TDG in the northern tail of Arp 245 was previously
studied in the optical, near-IR, HI, and CO by
\citet{duc00}.
This source is gas-rich, with an HI mass of 9 $\times$ 10$^8$ M$_{\sun}$.
Young stars are present in this feature, however,
the starlight is dominated by light from older stars, with the burst
contributing just a small fraction of the observed optical light
\citep{duc00}.
This candidate TDG has possible signs of rotation, 
however, this is uncertain
\citep{duc00}.
It has an oxygen abundance log(O/H) + 12 $\sim$ 8.6 $-$ 8.7 \citep{duc00}.
In the GALEX maps (Figure 16), 
the northern tail is bifurcated, as in the Arp 244
tail, with the two strands intersecting near the candidate TDG.
A UV-bright foreground star lies in front of
the northern tail; this was excluded in the regions used to get
the GALEX fluxes.

\noindent{\bf Arp 253}: Arp 253 (Figure 16)
is a pair of equal mass edge-on disk galaxies
seen end-to-end.  No strong tails are visible in this system.

\noindent{\bf Arp 254}: Arp 254 (Figure 16) is a pair of equal mass
disk galaxies, with the northern galaxy seen face-on, and the southern
edge-on.   The southern galaxy has short tidal tails.

\noindent{\bf Arp 261:} 
Arp 261 (Figure 17) may be an example of a `Taffy' galaxy,
produced by a direct head-on collision
between two equal-mass gas-rich galaxies.
During such collisions,
the impact may be sufficient to ionize the gas and strip it
from the disks, leaving a large quantity of gas between
the two galaxies \citep{condon93}.  
In Taffy galaxies, the bridge is bright in radio
synchrotron emission due to the magnetic field being stripped
from the galaxy along with the interstellar matter (Condon et al.\ 1993).
Only two Taffy systems
have been identified to date, UGC 813/6 and UGC 12914/5 
\citep{condon93, condon02},
thus the possible existence of a third in the nearby
Universe is of great interest.

Arp 261 is a close pair of edge-on irregular or spiral galaxies, 
with an optical and UV morphology suggesting a recent head-on collision.  
The western galaxy in the pair has a possible ring-like structure,
while an apparent bridge 
is visible to the northeast
between the two galaxies. 
In the NRAO VLA Sky Survey (NVSS), the radio continuum map
shows a bright peak between the two galaxies, suggesting
a radio continuum bridge between the
two galaxies, as in UGC 12914/5 and UGC 813/6. This
needs to be confirmed with higher resolution images.
Arp 261 is much closer than the other two Taffy galaxies,
at only 27 Mpc, and has a lower B luminosity,
suggesting that it may consist of two lower mass dwarf galaxies
rather than spirals.
In Figure 17,
a third galaxy, KTS 52, 
is visible to the northeast
of the pair.  This has the same redshift as the other two galaxies,
thus is part of the same system.

\noindent{\bf Arp 269:} 
Arp 269 (Figure 10) is an unequal-mass pair of galaxies connected by
a bridge.   In both the GALEX and SDSS images,
an off-center group of blue star forming regions is visible
in the smaller galaxy NGC 4485, as well as
along the bridge.   
As noted by \citet{elmegreen98}, several of these knots of
star formation lie in a tail-like structure extending to the
southwest of NGC 4485, towards the bridge.
These star formation
knots in the bridge and NGC 4485
are also seen in archival Spitzer infrared images.

In the GALEX images, a short (2$'$) 
tail-like feature is visible extending
to the east of NGC 4490, and some diffuse UV emission
is seen to the north of NGC 4490. 
In the \citet{clemens98a} 
HI maps, two large plume-like features
are seen extending 10$'$ ($\sim$20 kpc)
to the north of NGC 4485 and to the south of the
larger galaxy NGC 4490.
Neither of these plumes are strongly
detected in the GALEX or SDSS images,
however, smoothed SDSS images show
a possible hint of the southern plume.
This lack of UV emission is surprising in light of
the relatively high HI column density in
the inner 5$'$ (10 kpc) sections of
these plumes of 4 $\times$ 10$^{20}$
cm$^{-2}$.

It was suggested by
\citet{clemens98a} that the HI plumes in Arp 269 were produced
by SN-driven outflow from the main galaxy NGC 4490.
Furthermore, 
\citet{clemens00} suggest that
the smaller galaxy NGC 4485 passed through
the gaseous
disk of the larger galaxy NGC 4490, and ram pressure from the
collision may have caused an offset in the location of the interstellar
gas in NGC 4485, and thus the observed offset in star formation.
We suggest an alternative possibility, that
the HI plumes are simply gas-rich tidal features, and the star formation
in NGC 4485 and along the bridge was triggered by
gas flowing
from the larger galaxy NGC 4490 to NGC 4485 along the bridge.
Thus this may be an example of accretion from one galaxy to another,
as in Arp 285.
Deeper optical and UV imaging is needed to search for a stellar component
to the HI plumes, to distinguish between outflow or tidal origin.

\noindent{\bf Arp 270:} Arp 270 (Figure 10)
is a close pair of equal-mass spirals with
a very long (10$'$ = 79 kpc) thin HI tail to the southeast \citep{clemens99}.
This tail has
typical HI column densities of $\sim$1.8 $\times$ 10$^{20}$ 
cm$^{-2}$, and a concentration near the end with 
N$_{HI}$ $\sim$ 4 $\times$ 10$^{20}$ cm$^{-2}$
\citep{clemens99}.
This HI concentration was classified as a possible tidal dwarf galaxy by 
\citet{clemens99}.
In the smoothed GALEX
images, there is faint diffuse emission near the middle of this tail.
In the HI concentration near the end 
and another HI clump 3$'$ from
the end, faint UV sources are visible in the GALEX and SDSS maps. 
The source near the end of the tail (labeled `TDG' in Table 3) has
moderate optical/UV colors, while the second source is 
quite blue in FUV $-$ NUV and NUV $-$ g.
These may be extended in the SDSS images.
Optical spectroscopy is needed to confirm that these sources are
associated with the tail.

The HI maps also show a shorter tail extending
to the north from the western side of the 
pair \citep{clemens99}.  The base of 
this tail is visible in the UV and optical
images, with two UV-bright knots evident in the GALEX images.

Further south of the Arp 270 pair, the dwarf galaxies IC 2604 and IC 2608
are visible on the GALEX images.  These are also present in
the HI maps of \citet{clemens99}, and are at the same redshift as Arp 270.

\noindent{\bf Arp 271:} 
Arp 271 (Figure 17)
is a close pair of grand design spirals seen almost face-on.
Between the two galaxies a double bridge is seen.
In the GALEX maps, 
the spiral patterns and the two bridge strands are
prominent.
The western bridge component has a UV-bright
knot which also is present in our SARA H$\alpha$ map,
as well as the H$\alpha$ maps of \citet{evans96}.
In the \citet{clemens98b} VLA HI map, the bridge between
the galaxies is visible, along with two faint
tail-like structures extending 4$'$ (30 kpc) to the west
from each of the main galaxies. 
These have relatively low HI surface brightnesses of 8 $\times$ 10$^{19}$
cm$^{-2}$, thus the lack of associated diffuse UV emission in these features
in the GALEX maps is not surprising.
Two UV-bright clumps are visible at the base of the southern tail.
There is also a UV point source in the northern tail, and
another UV source in a faint HI clump to the south of the southern tail.
Whether these UV sources are at the same redshift
as Arp 271 is unknown. 
In the \citet{clemens98b} HI map, a dwarf galaxy named
APMUKS
B1400004.67-054820.9 is seen 8$'$ (60 kpc) 
to the west of the pair, beyond the end of the tails.
This galaxy is bright in the GALEX maps.

\noindent{\bf Arp 280:} 
Arp 280 (Figure 11)
is an edge-on disk galaxy (SB(r)b: in NED) with a small irregularly-shaped
companion (SBm pec in NED).   Neither galaxy has strong tails in the optical
or UV.  The HI map of \citet{clemens98b} shows a short ($\sim$1$'$ = 4 kpc)
tail to the southeast and a longer ($\sim$4$'$ = 16 kpc) tail to the northwest.
These tails are not seen in the GALEX or SDSS images, in spite of the 3$-$4 
$\times$ 10$^{20}$ cm$^{-2}$ HI column density.

\noindent{\bf Arp 282:}
Arp 282 (Figure 11) may be a collisional ring galaxy; alternatively, it
may be a spiral gravitationally disturbed by a smaller companion.
It consists of two edge-on disk galaxies, orientated such that
their major axes lie perpendicular to each other, with the smaller
galaxy being near the major axis of the larger galaxy.
The larger galaxy has disturbed spiral arms or ring-like structures
extending to the east in the SDSS images.
These features are called the `eastern tail' in Table 3,
but they may be rings or spiral arms
rather than a classical tidal tail.
Two bright knots of star formation are
visible in these arcs.   
The two galaxies are
connected by a faint bridge, and 
a short plume extends to the south of the smaller galaxy.
A bright star to the east of this pair may
affect the measured photometry.

\noindent{\bf Arp 283:} 
Arp 283 (Figure 12) is an equal-mass pair of spirals, neither of which has long tidal
features.   The eastern galaxy in the pair, NGC 2799, is viewed edge-on, and the portion
of the disk closest to the companion (called the `bridge' in Table 3) is tidally
disturbed, with star forming regions visible in the GALEX, SDSS, Spitzer,
and SARA H$\alpha$ maps.
In the SDSS images, the western galaxy NGC 2798 has two broad smooth arms/tails
(called `north tail' and `south tail' in Table 3)
without obvious clumps.  These are less prominent in the GALEX data than
in the optical.   As noted earlier, these have very red FUV $-$ NUV colors.

\noindent{\bf Arp 284:} 
In the \citet{arp66} Atlas picture of Arp 284, the western galaxy NGC 7714
shows a partial ring, two tails to the west, another tail to the northeast,
and is connected to its edge-on companion NGC 7715 by a bridge.
Strong star formation is visible in the NGC 7714 disk as well as
in the bridge \citep{smith97}.
In HI maps, the outer western tail seen in the optical images
loops back to the bridge, and another long tail is visible to the far west
\citep{smith92, smith97}.
The basic morphology of this system can be reproduced by a prograde,
near-head-on collision \citep{struck03}, with the inner western tail
being formed via accretion from the bridge, the HI loop being a classical
tidal feature, and the far west HI tail being the end of the eastern tail
of NGC 7715, wrapped behind the galaxies.
In the GALEX FUV map (Figure 17), the 
HI loop is detected, while the inner `accretion tail'
is very bright.  The section of the HI loop that contains a possible
`ultraluminous X-ray' (ULX) point source \citep{smith05b} is particularly
bright in the UV.
The knots of star formation in the bridge are also quite
prominent in the UV.
In the far west HI tail \citep{smith92}, a hint of extended emission is
seen in the smoothed FUV map.
A detailed analysis of the GALEX data,
in conjunction with Spitzer and Hubble images, 
has been conducted by \citet{peterson09}.

\noindent{\bf Arp 285:} 
The northern galaxy in Arp 285 (Figure 12), 
NGC 2856, has a peculiar tail-like structure
extending to the north perpendicular to the disk.
This is very blue in the GALEX/SDSS colors, being the bluest tail in our
sample in g $-$ r.
According to our numerical simulations of this system, this feature
was produced from material accreted along the bridge from the companion
\citep{smith08}.   
The southern galaxy in this system, NGC 2854, has a curved tail extending
to the south in the GALEX data, matching the HI tail of \citet{chengalur94, chengalur95}.
The GALEX and SDSS images of this system have been discussed in detail
in \citet{smith08}.   

\noindent{\bf Arp 290:}
Arp 290 (Figure 13) contains two unequal mass disk galaxies.
The larger more northern galaxy
has an apparent bar and two peculiar UV-bright arm-like
features.  As noted above, the smaller southern galaxy 
IC 195 is classified as SAB0 and is very red in FUV $-$ NUV.

\noindent{\bf Arp 295:} Arp 295 (Figure 17) is a wide pair of
equal mass disk galaxies with tails and a long bridge. 
It was included in the HI and optical imaging study of 
\citet{hibbard96}.

\noindent{\bf Arp 297N:} 
Arp 297 (Figure 13)
consists of two unrelated pairs of galaxies at different redshifts.
The more northern pair includes NGC 5755, 
a disturbed
spiral with prominent tails/arms, and 
NGC 5753, a more compact galaxy. 

\noindent{\bf Arp 297S:} 
In the Arp Atlas image, the 
southern pair in Arp 297 resembles M51, 
with 
a face-on grand design spiral
NGC 5754
and 
a smaller compact galaxy
NGC 5752.
However, 
in the smoothed GALEX and SDSS images 
a 5$'$ (90 kpc) long faint tail extending 
to the west is visible
(Figure 33).
This feature is also 
visible 
in deep optical
images \citep{keel03}.

\noindent{\bf Arp 305:} 
A detailed study of the GALEX and SDSS
data for the widely separated pair of spirals in Arp 305
has been published in \citet{hancock09}.
In these images (Figure 14), a luminous collection of young stars
is seen between the two galaxies.   This is located in a large
concentration of atomic gas within a gaseous bridge, as seen
in HI maps \citep{vanmoorsel83}.  The inferred stellar mass
of this bridge source is 1$-$7 $\times$ 10$^6$ M$_{\sun}$ \citep{hancock09}, 
consistent with it being a low mass tidal dwarf galaxy.
Significant tidal debris is also seen around the southern
galaxy NGC 4017 in the GALEX maps.  
According to our numerical simulation of this encounter,
the bridge TDG will eventually fall into the primary galaxies,
and will not become an independent dwarf galaxy.

\noindent{\bf NGC 4567:} NGC 4567 (Figure 14) is a disturbed
pair of spirals without strong tidal tails.

\section{Summary}

In this study, we presented GALEX UV images of 42 interacting galaxy
pairs, and compared with available data at other wavelengths.  
In these images, 
we identified numerous examples of `beads on a string' and accretion
from one galaxy to another, as well as a few new possible tidal dwarf 
galaxies and ring galaxies.  
We also note a possible `Taffy' galaxy in our sample.
We compared the distributions of UV $-$ optical colors
of the tidal features of our sample galaxies with those
of the main disks, and found little difference.  However, when
comparing tidal features with their own parent galaxies
on average they appear somewhat bluer in NUV $-$ g.   This effect
may be due to recent star formation and/or lower metallicities on average.

This dataset
provides a valuable atlas of the rest frame UV morphology for a large
set of nearby interacting galaxies with a variety of structures.  This
dataset will be 
very useful for interpreting the 
morphology of distant galaxies.
Such studies have recently been done for smaller samples of
nearby interacting galaxies,
which have been artificially redshifted to high redshifts
to compare with distant galaxies
(e.g., \citealp{overzier08, petty09}).
At high redshifts, the merger rate increases and 
galaxies appear more perturbed (i.e., \citealp{abraham96, lotz06}).
At z $\sim$ 2 deep optical ground-based surveys 
trace the rest frame UV with a resolution of 
$\sim$0.5$''$ ($\sim$4.2 kpc at z = 2). 
The GALEX beam corresponds to 
scales of 0.6 $-$ 3.6 kpc for our sample, which nicely matches the best 
resolution obtained for z $\sim$ 2 systems.  

In a follow-up paper, we will
compare the UV/optical colors of our interacting galaxies with those
of a `control' sample of normal spirals, to search for
evidence of enhancement in the star formation rate relative to
less disturbed systems.
Additional
planned 
work includes a detailed comparison of the GALEX
and optical images with available HI, H$\alpha$,
and Spitzer maps, to investigate
star formation efficiencies and thresholds 
in the tidal features.
We also plan
detailed analyses of the spatial distribution of star formation
within selected galaxies in the sample.
We have acquired ground-based optical spectra and 
Spitzer infrared spectra of a few
of the star forming regions in our sample, which will be
presented in future papers \citep{hancock10, higdon10}.





\acknowledgments

We thank the GALEX team for making this research possible.
This research was supported by GALEX grant GALEXGI04-0000-0026,
NASA LTSA grant NAG5-13079, and Spitzer
grants RSA 1353814 and RSA 1379558.
This research has made use of the NASA/IPAC Extragalactic Database (NED) 
and the NASA/ IPAC Infrared Science Archive, 
which are operated by the Jet Propulsion Laboratory, 
California Institute of Technology, under contract with the National Aeronautics and Space Administration. 
We also utilized the GOLDMINE database (http://goldmine.mib.infn.it).

\clearpage




\begin{figure}
\caption{
  \small 
A comparison of the GALEX images (left) and the SDSS images (right).
North is up and east to the left in these figures, as well as in
all the subsequent UV and optical images.
When both FUV and NUV data are available, 
the displayed GALEX images are color-coded such that blue is the FUV data,
while yellow is the NUV.
For the SDSS pictures, the images are approximately true-color.
In the Arp 24 SDSS image, the small galaxy in the upper left
is a background galaxy.
}
\end{figure}

\begin{figure}
\caption{
  \small 
A comparison of the GALEX images (left) and the SDSS images (right), as
in Figure 1.
}
\end{figure}

\begin{figure}
\caption{
  \small 
A comparison of the GALEX images (left) and the SDSS images (right), as
in Figure 1.
The SDSS image of Arp 85 (M51) is a mosaic from \citet{hogg07}.
}
\end{figure}

\begin{figure}
\caption{
  \small 
A comparison of the GALEX images (left) and the SDSS images (right), as
in Figure 1.
In the Arp 87 image, no redshift is
available for the third galaxy to the upper left.
}
\end{figure}

\begin{figure}
\caption{
  \small 
A comparison of the GALEX images (left) and the SDSS images (right), as
in Figure 1.
The SDSS image of Arp 101 is a mosaic from \citet{hogg07}.
In the GALEX image of Arp 101, the brightest UV source near
the top of the Figure is not part of Arp 101.
}
\end{figure}

\begin{figure}
\caption{
  \small 
A comparison of the GALEX images (left) and the SDSS images (right), as
in Figure 1.
}
\end{figure}

\begin{figure}
\caption{
  \small 
A comparison of the GALEX images (left) and the SDSS images (right), as
in Figure 1.
}
\end{figure}

\begin{figure}
\caption{
  \small 
A comparison of the GALEX images (left) and the SDSS images (right), as
in Figure 1.
}
\end{figure}
\clearpage

\begin{figure}
\caption{
  \small 
A comparison of the GALEX images (left) and the SDSS images (right), as
in Figure 1.
}
\end{figure}

\begin{figure}
\caption{
  \small 
A comparison of the GALEX images (left) and the SDSS images (right), as
in Figure 1.
The SDSS image of Arp 270 is a mosaic from \citet{hogg07}.
}
\end{figure}

\begin{figure}
\caption{
  \small 
A comparison of the GALEX images (left) and the SDSS images (right), as
in Figure 1.
A bright star lies to the northeast of Arp 282, beyond
the displayed field of view.
}
\end{figure}

\begin{figure}
\caption{
  \small 
A comparison of the GALEX images (left) and the SDSS images (right), as
in Figure 1.
The SDSS image of Arp 285 is a mosaic from \citet{hogg07}.
}
\end{figure}

\begin{figure}
\caption{
  \small 
A comparison of the GALEX images (left) and the SDSS images (right), as
in Figure 1.
For the Arp 297 images,
to the west of the southern pair beyond the extent of this
view, a long tail is visible in
the smoothed SDSS and GALEX images.
}
\end{figure}

\begin{figure}
\caption{
  \small 
A comparison of the GALEX images (left) and the SDSS images (right), as
in Figure 1.
}
\end{figure}

\begin{figure}
\caption{
  \small 
The GALEX images for the galaxies without available SDSS data.
These images are color-coded such that blue is the FUV data,
while yellow is the NUV.
In the Arp 65 image,
the small angular size galaxy to the northeast is a background galaxy. 
}
\end{figure}

\begin{figure}
\caption{
  \small 
The GALEX images for the galaxies without available SDSS data.
These images are color-coded such that blue is the FUV data,
while yellow is the NUV.
}
\end{figure}

\begin{figure}
\caption{
  \small 
GALEX pictures of sample galaxies
without SDSS observations.
}
\end{figure}

\begin{figure}
\caption{
  \small 
GALEX pictures of sample galaxies
without SDSS observations.
}
\end{figure}

\begin{figure}
\plotone{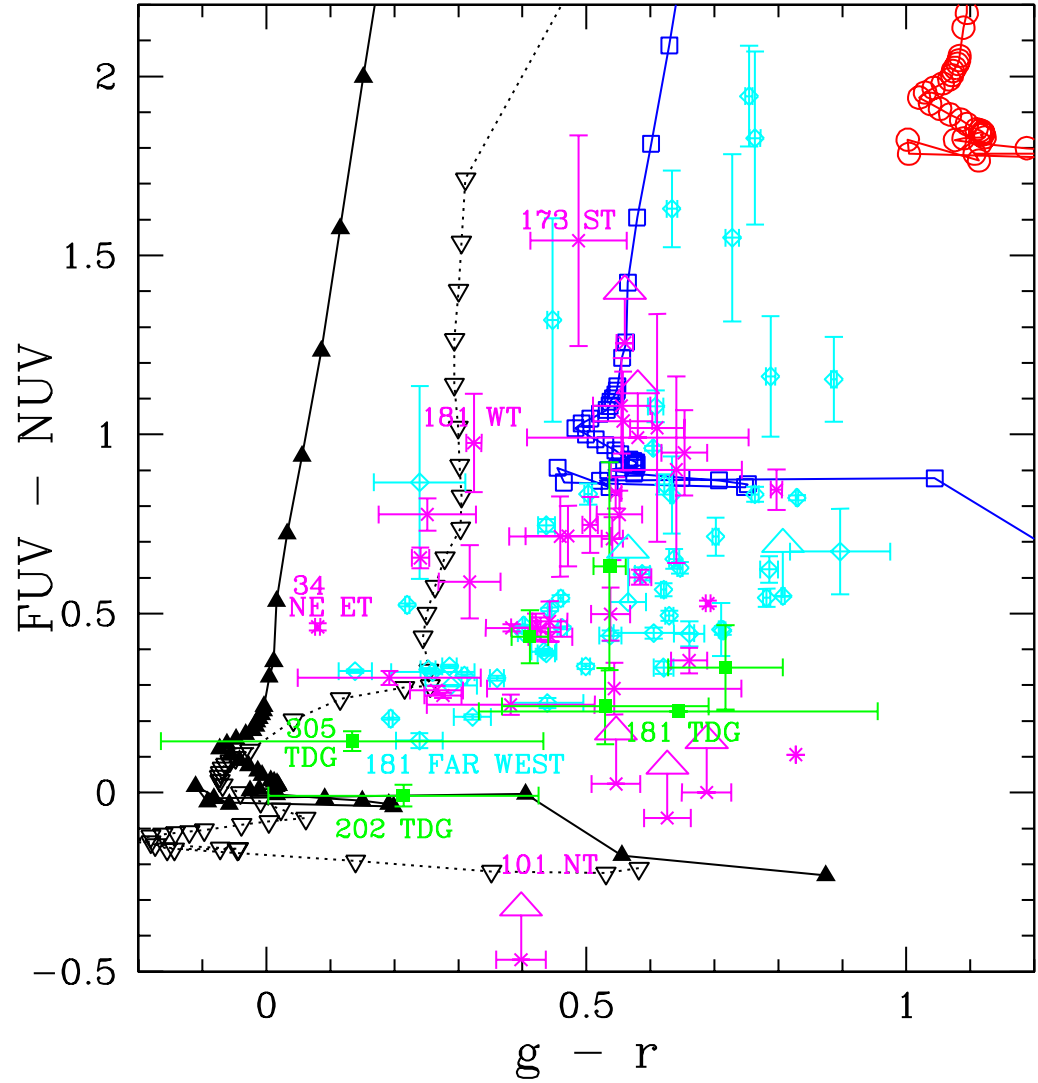}
\caption{
  \small 
The GALEX FUV $-$ NUV vs.\ g $-$ r colors for the 
main disks of the interacting galaxy sample (cyan open diamonds)
and the tidal features of these galaxies (magenta crosses).
The green filled squares are the values for the candidate TDGs, which have been
measured separately from their parent tidal features.
Some of these features are labeled, in the same color as the corresponding
data point.
The solar metallicity models have an extinction of E(B $-$ V) = 0 
(black filled triangles),
E(B $-$ V) = 0.5 (blue open squares), and E(B $-$ V) = 1 (open red circles).
The upside down black open triangles are zero extinction 0.2
solar metallicity models.
All models include H$\alpha$.
The ages are increasing from the bottom starting at 1 Myrs, by step sizes of 1 Myrs to 20 Myrs,
then by 5 Myr steps to 50 Myrs, then 10 Myr steps to 100 Myrs, 100 Myr
steps to 1 Gyr, and 500 Myr steps to 10 Gyrs.
}
\end{figure}

\clearpage

 \begin{figure}
\plotone{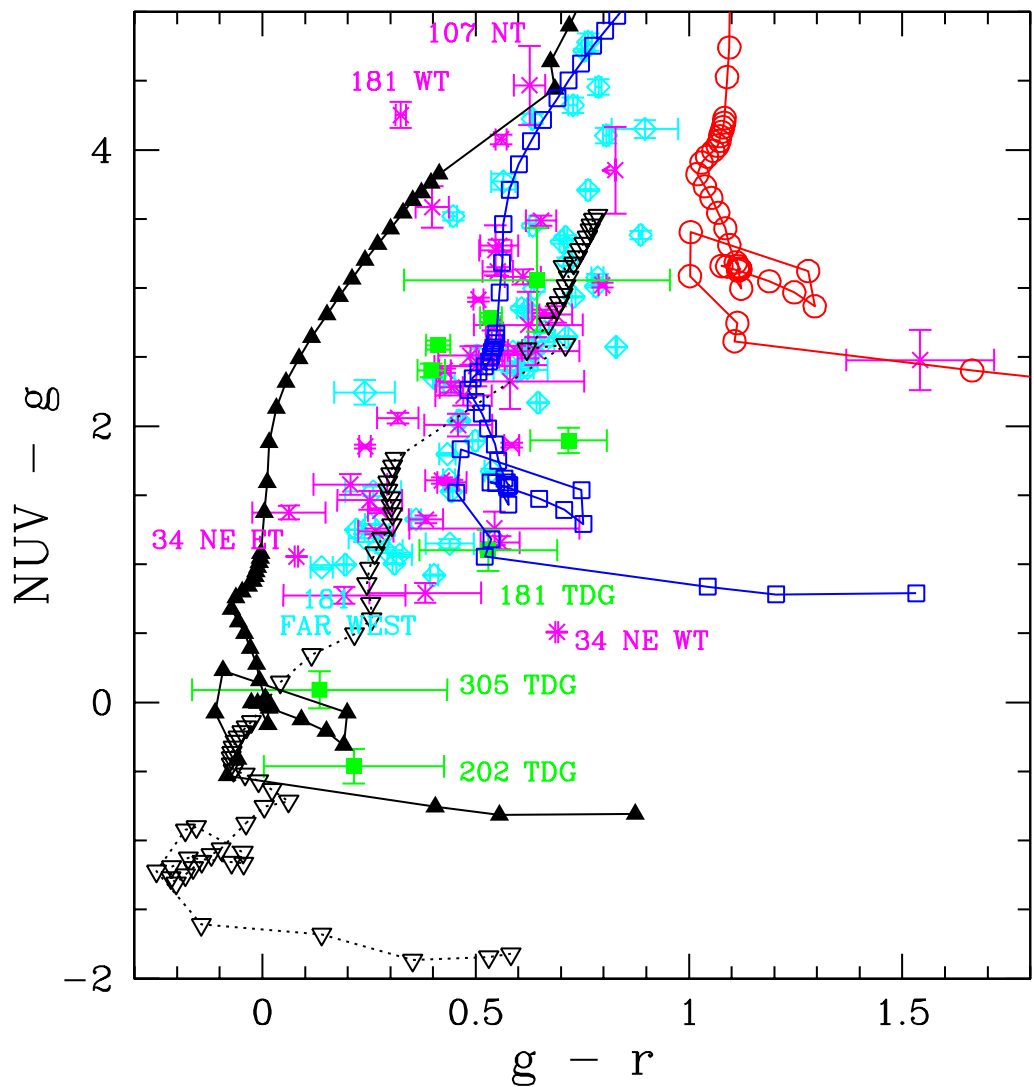}
\caption{
  \small 
The GALEX/SDSS NUV $-$ g vs.\ g $-$ r colors for the 
main disks of the interacting galaxy sample (cyan open diamonds)
and the tidal features of these galaxies (magenta crosses).
The green filled squares are the values for the candidate TDGs, which have been
measured separately from their parent tidal features.
The solar metallicity
models have an extinction of E(B $-$ V) = 0 (black filled triangles),
E(B $-$ V) = 0.5 (blue open squares), and E(B $-$ V) = 1 (open red circles).
The upside down black open triangles are 0.2 zero extinction
solar metallicity models.
All models include H$\alpha$.
The ages are increasing from the bottom starting at 1 Myrs, 
by step sizes of 1 Myrs to 20 Myrs,
then by 5 Myr steps to 50 Myrs, then 10 Myr steps to 100 Myrs, 100 Myr
steps to 1 Gyr, and 500 Myr steps to 10 Gyrs.
}
\end{figure}

 \begin{figure}
\plotone{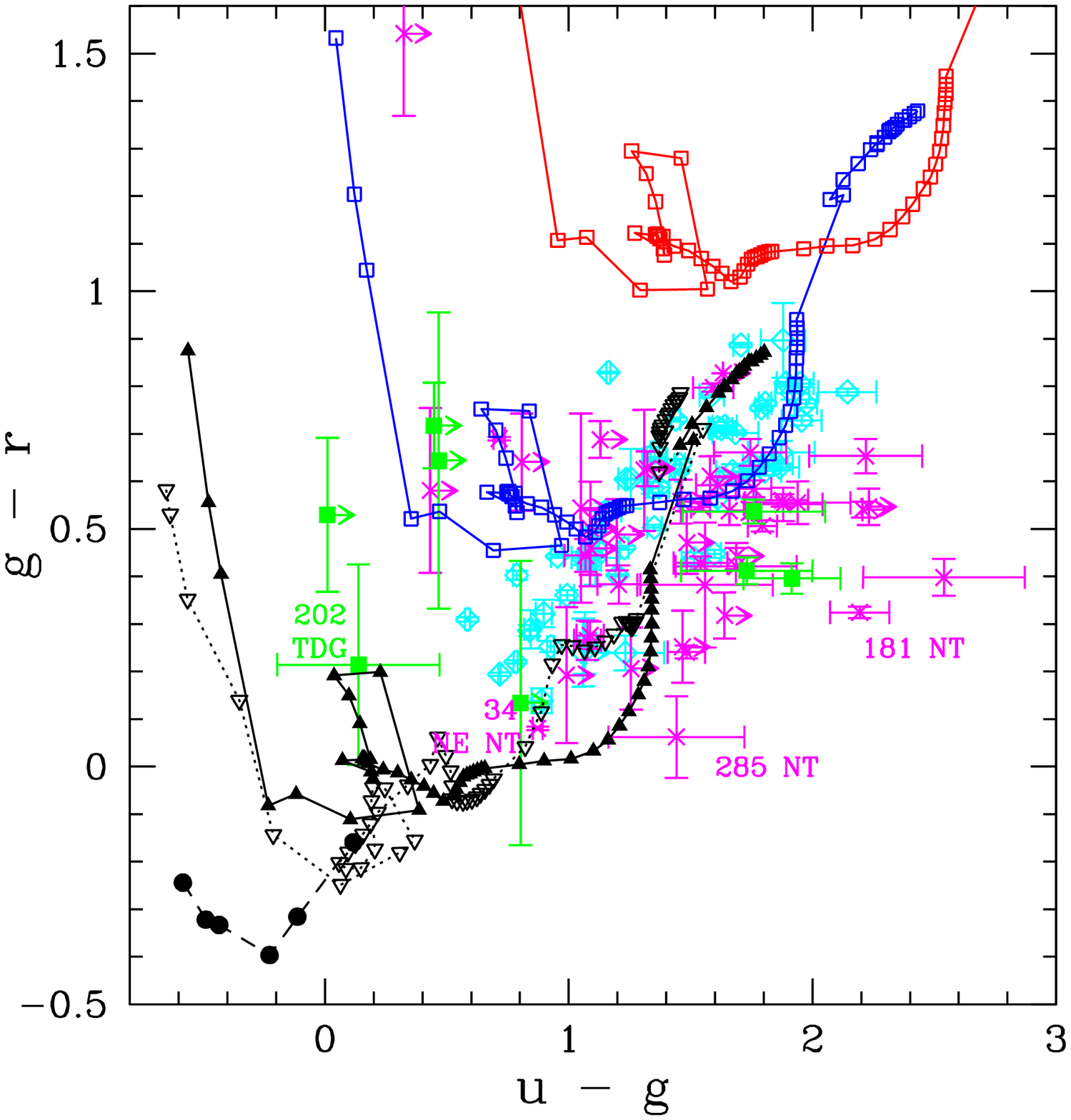}
\caption{
  \small 
The SDSS g $-$ r vs.\ u $-$ g colors for the 
main disks of the interacting galaxy sample (cyan open diamonds)
and the tidal features of these galaxies (magenta crosses).
The green filled squares are the values for the candidate TDGs, which have been
measured separately from their parent tidal features.
The models have an extinction of E(B $-$ V) = 0 (black filled triangles),
E(B $-$ V) = 0.5 (blue open squares), and E(B $-$ V) = 1 (open red circles).
The upside down black open triangles are 0.2 zero extinction
solar metallicity models.
All models include H$\alpha$, except for the filled black circles,
which are zero extinction solar metallicity models without H$\alpha$.
The ages are increasing from the bottom, by step sizes of 1 Myrs to 20 Myrs,
then by 5 Myr steps to 50 Myrs, then 10 Myr steps to 100 Myrs, 100 Myr
steps to 1 Gyr, and 500 Myr steps to 10 Gyrs.
}
\end{figure}

 \begin{figure}
\plotone{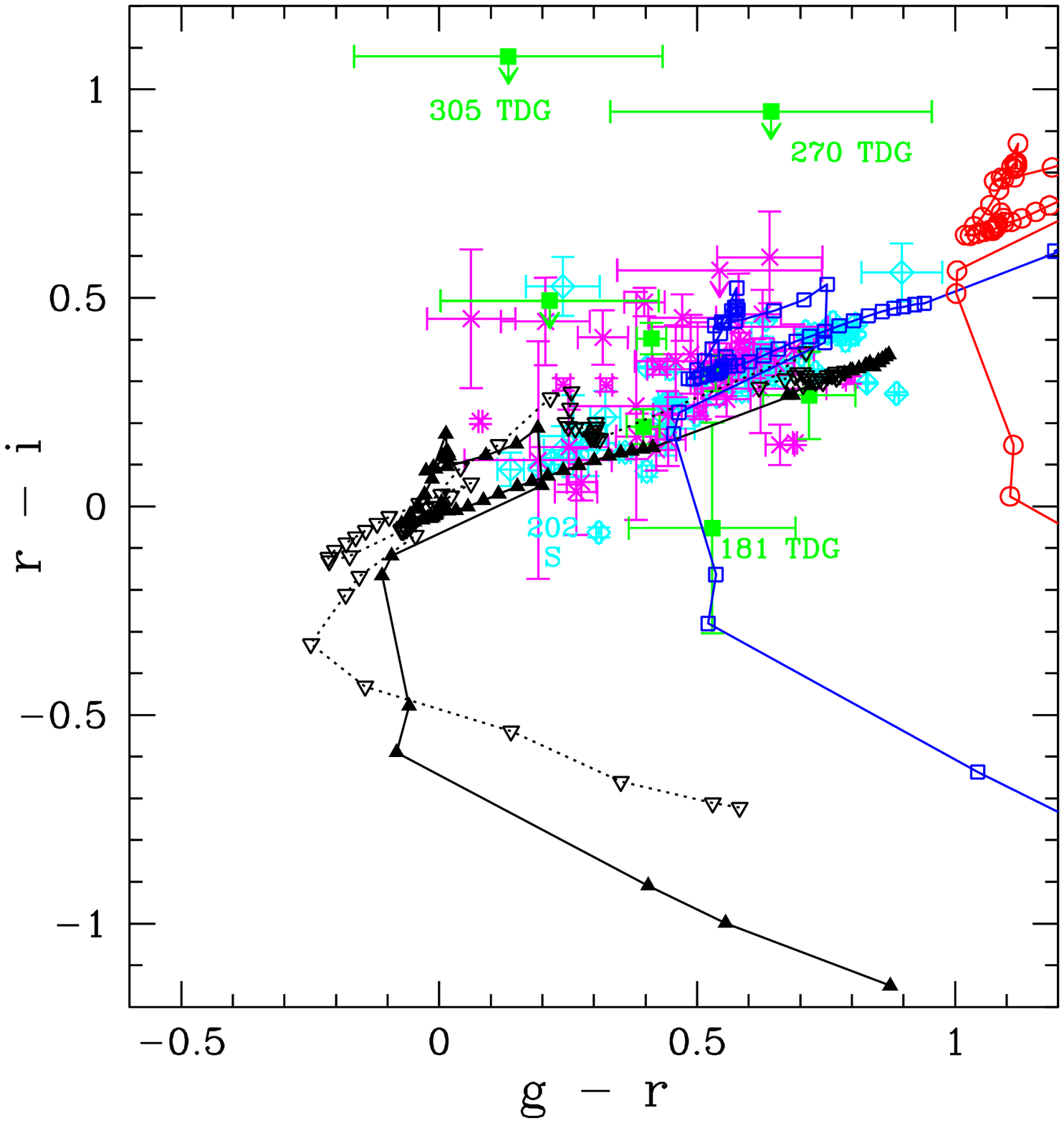}
\caption{
  \small 
The SDSS r $-$ i vs.\ g $-$ r colors for the 
main disks of the interacting galaxy sample (cyan open diamonds)
and the tidal features of these galaxies (magenta crosses).
The green filled squares are the values for the candidate TDGs, which have been
measured separately from their parent tidal features.
The models have an extinction of E(B $-$ V) = 0 (black filled triangles),
E(B $-$ V) = 0.5 (blue open squares), and E(B $-$ V) = 1 (open red circles).
The upside down black open triangles are 0.2 zero extinction
solar metallicity models.
All models include H$\alpha$.
The ages are increasing from the bottom, by step sizes of 1 Myrs to 20 Myrs,
then by 5 Myr steps to 50 Myrs, then 10 Myr steps to 100 Myrs, 100 Myr
steps to 1 Gyr, and 500 Myr steps to 10 Gyrs.
}
\end{figure}

 \begin{figure}
\plotone{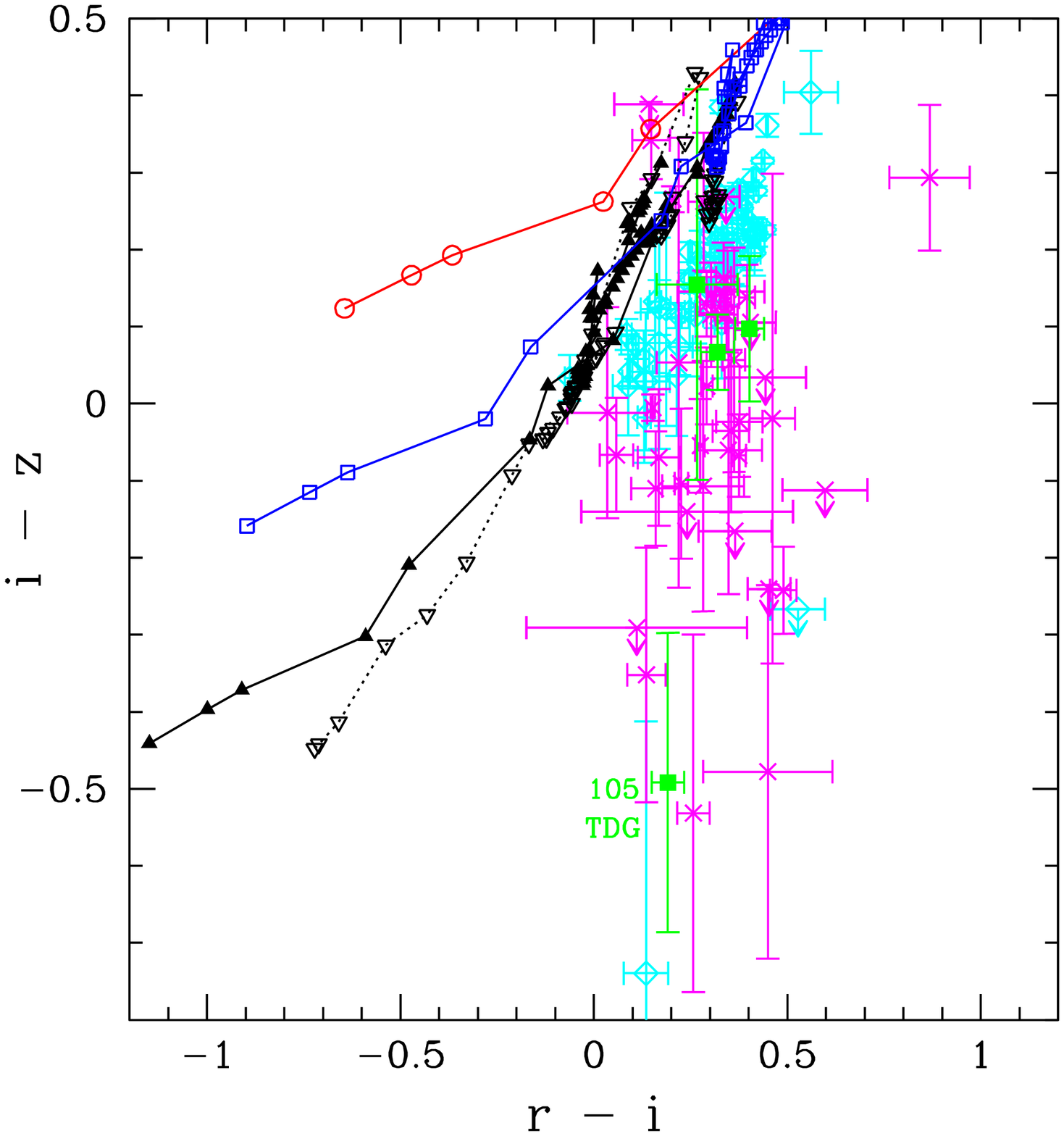}
\caption{
  \small 
The SDSS 
i $-$ z 
vs.\ 
r $-$ i 
colors for the 
main disks of the interacting galaxy sample (cyan open diamonds)
and the tidal features of these galaxies (magenta crosses).
The green filled squares are the values for the candidate TDGs, which have been
measured separately from their parent tidal features.
The models have an extinction of E(B $-$ V) = 0 (black filled triangles),
E(B $-$ V) = 0.5 (blue open squares), and E(B $-$ V) = 1 (open red circles).
The upside down black open triangles are 0.2 zero extinction
solar metallicity models.
All models include H$\alpha$.
The ages are increasing from the bottom starting at 1 Myrs, by step sizes of 1 Myrs to 20 Myrs,
then by 5 Myr steps to 50 Myrs, then 10 Myr steps to 100 Myrs, 100 Myr
steps to 1 Gyr, and 500 Myr steps to 10 Gyrs.
}
\end{figure}

\begin{figure}
\plotone{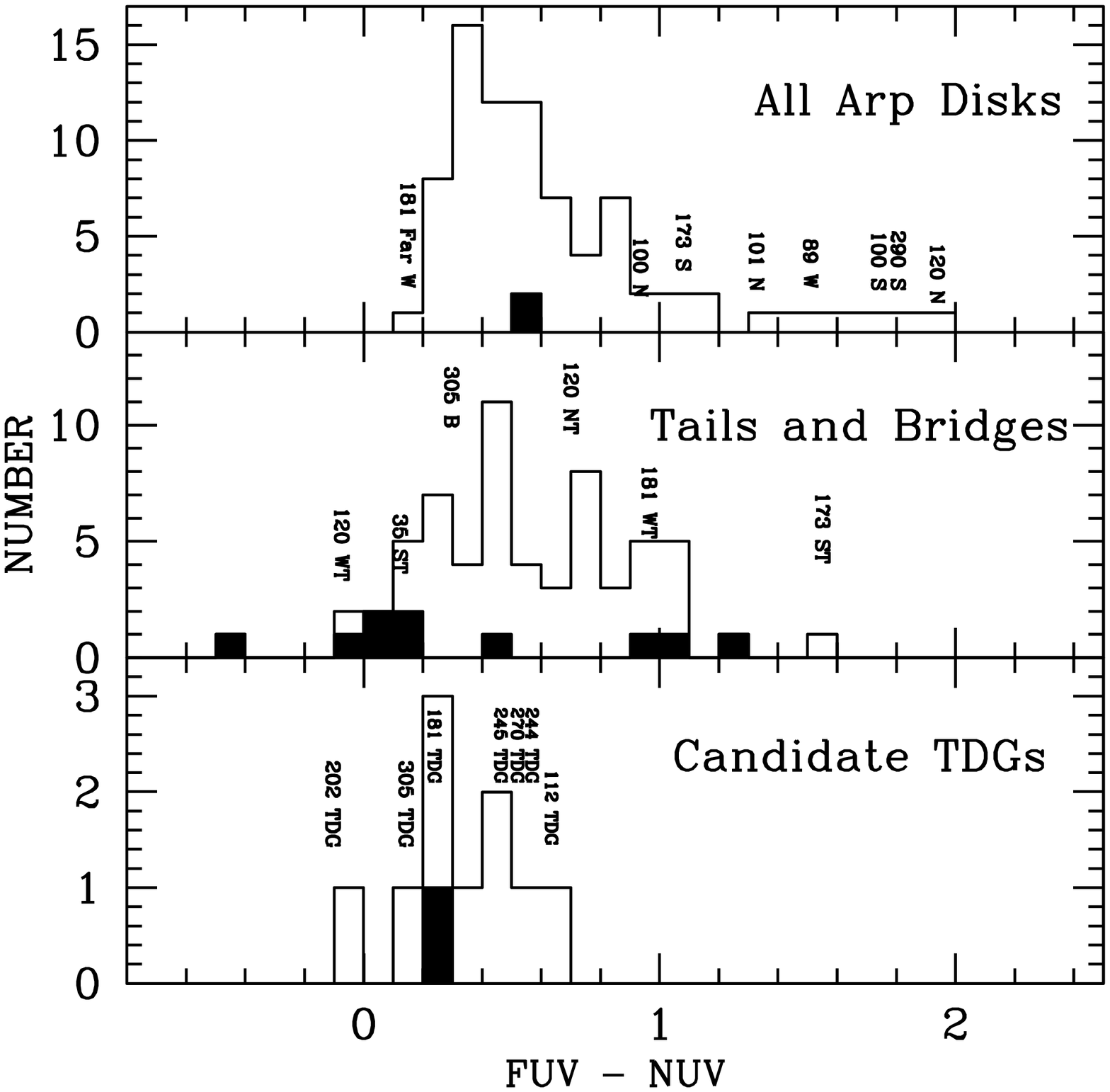}
\caption{
  \small 
Histogram of the FUV $-$ NUV colors of the Arp disks, 
the Arp tails and bridges, and the TDGs.
The filled regions are lower limits.
}
\end{figure}

\begin{figure}
\plotone{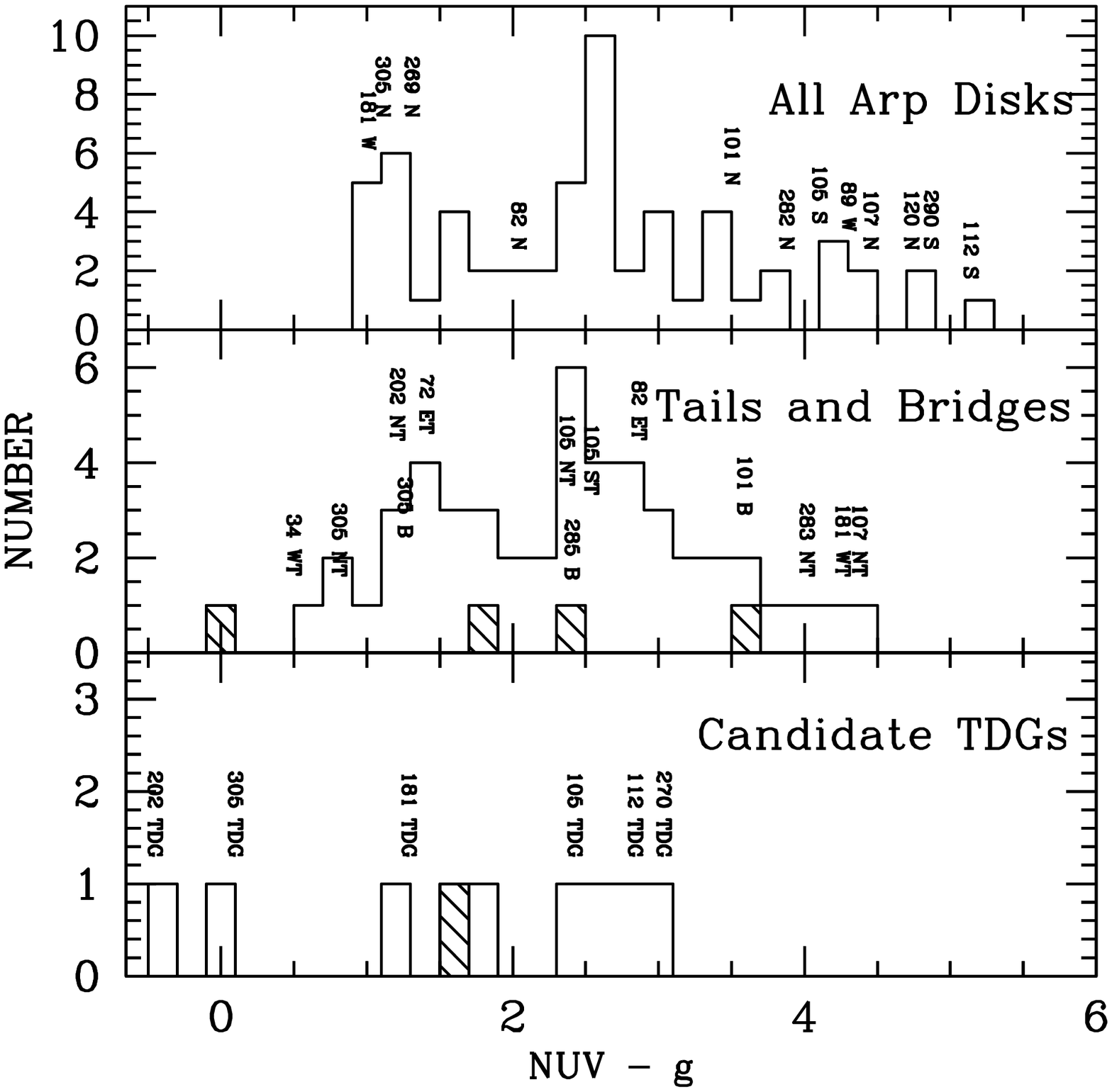}
\caption{
  \small 
Histogram of the NUV $-$ g colors of the Arp disks, the Arp tails and bridges,
and the TDGs.
The hatched areas are upper limits.
}
\end{figure}

\begin{figure}
\plotone{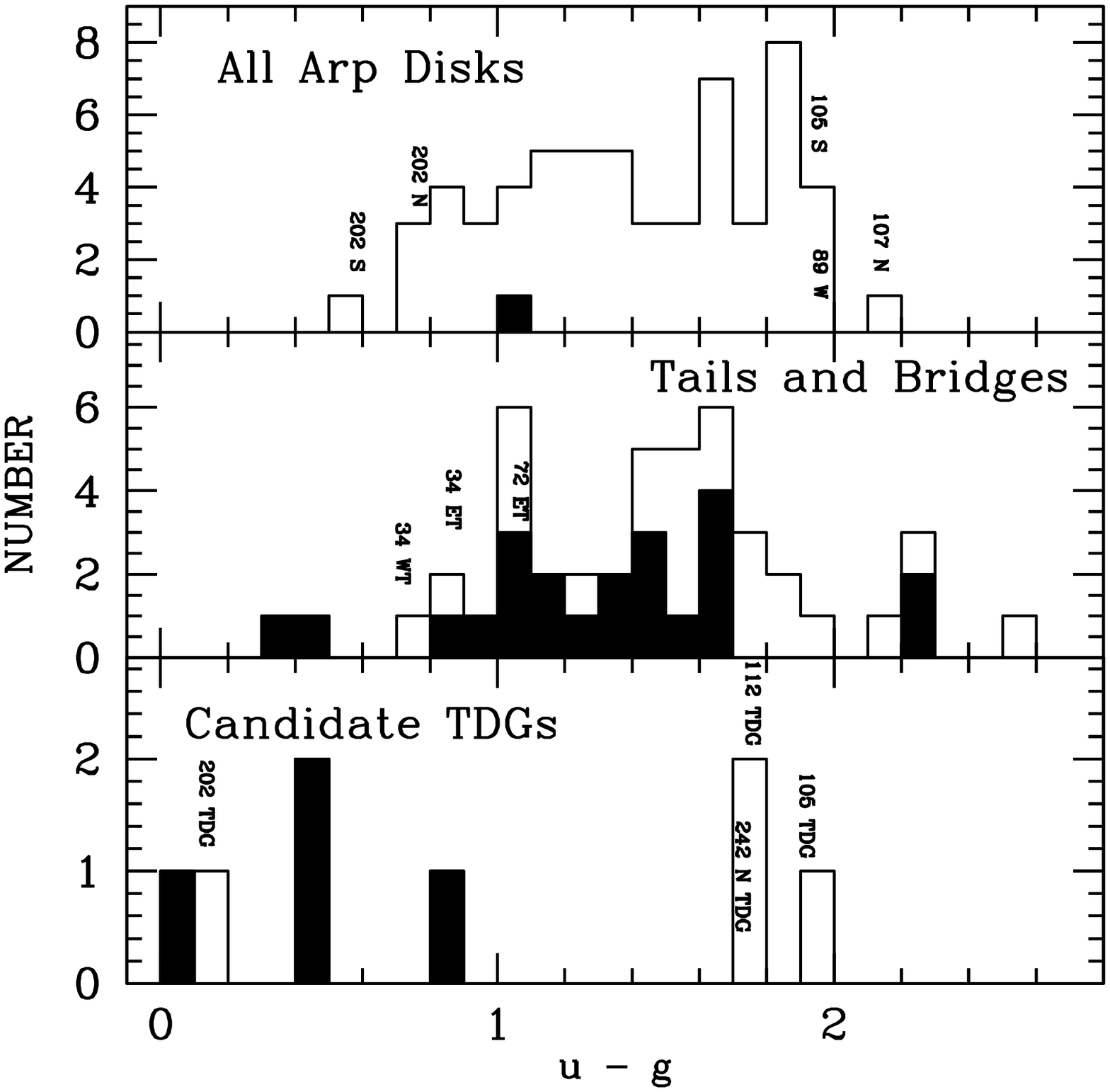}
\caption{
  \small 
Histogram of the u $-$ g colors of the Arp disks, the Arp tails and bridges,
and the TDGs.
The filled regions are lower limits.
}
\end{figure}

\begin{figure}
\plotone{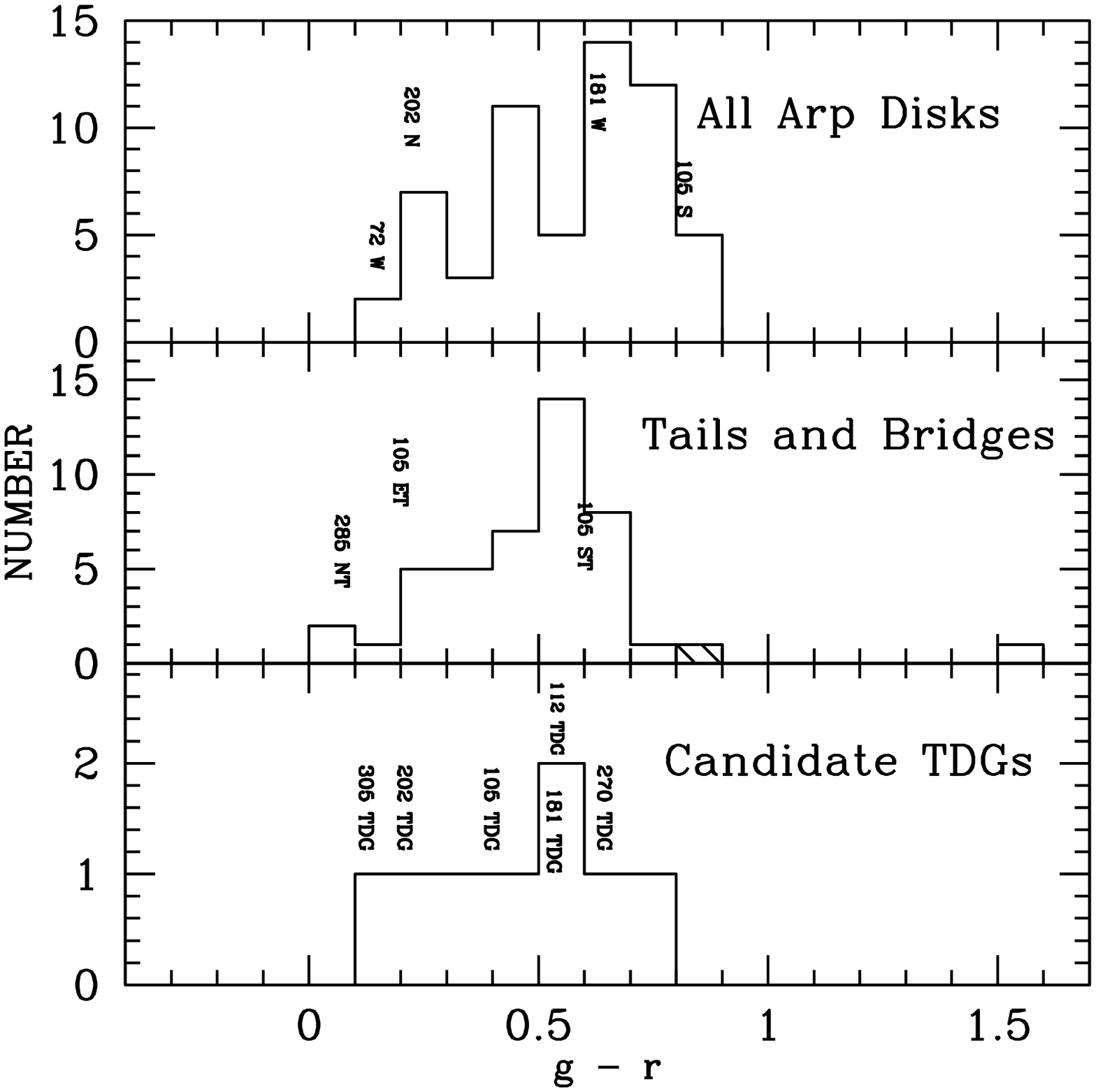}
\caption{
  \small 
Histogram of the g $-$ r colors of the Arp disks, the Arp tails and bridges,
and the TDGs.
The hatched regions are upper limits.
}
\end{figure}

\begin{figure}
\plotone{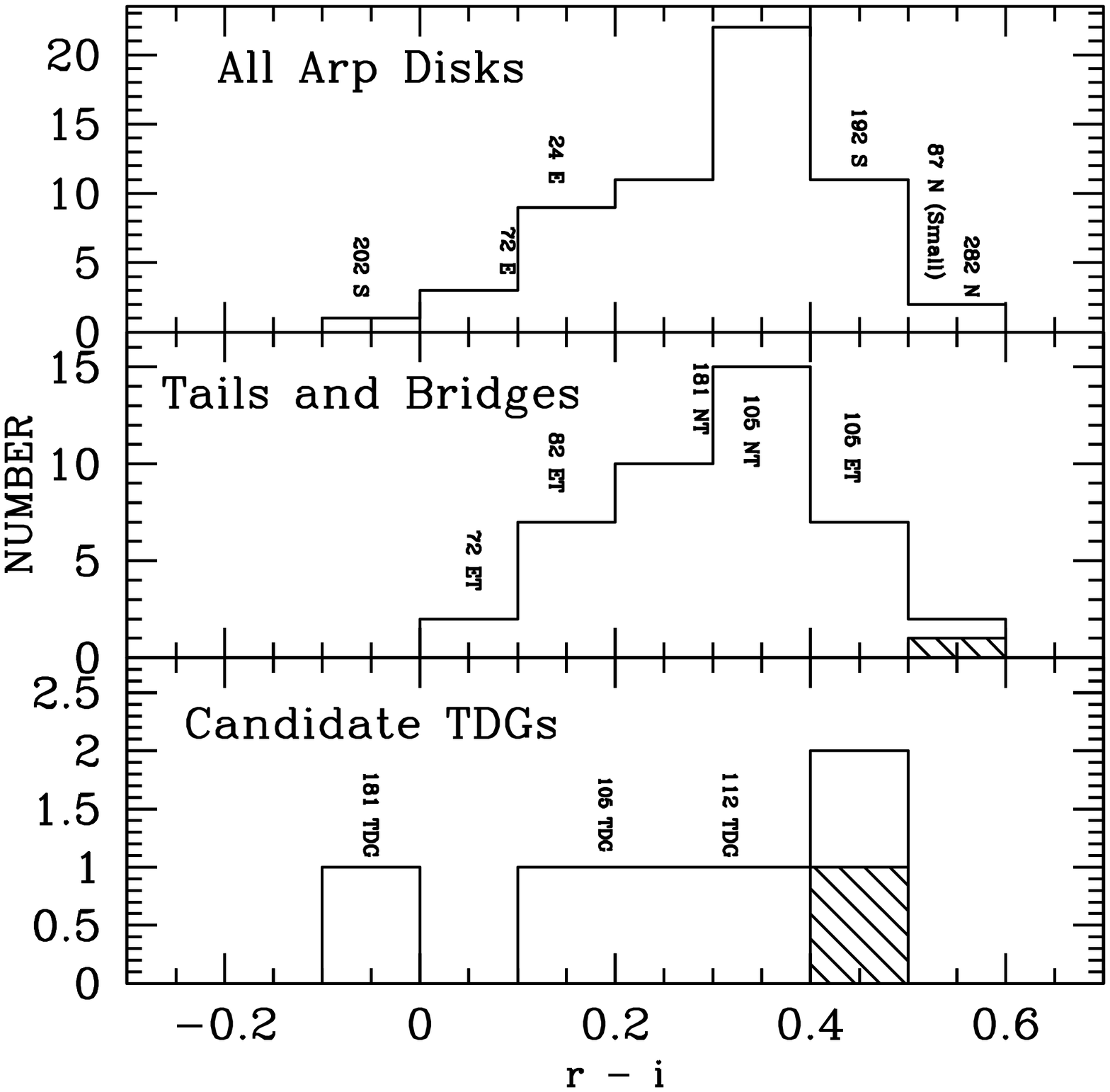}
\caption{
  \small 
Histogram of the r $-$ i colors of the Arp disks, the Arp tails and bridges,
and the TDGs.
The hatched areas are upper limits. 
}
\end{figure}

\begin{figure}
\plotone{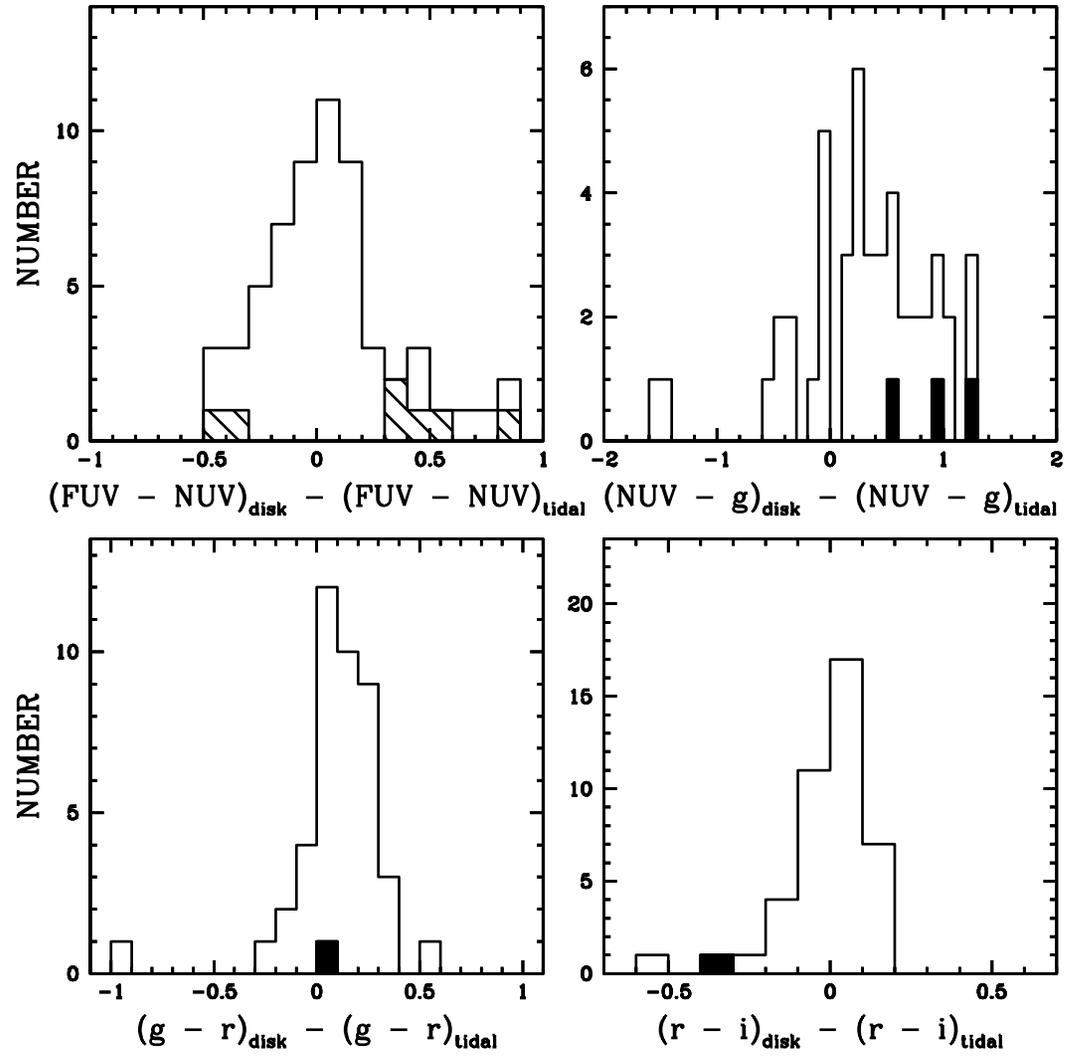}
\caption{
  \small 
Histograms of the difference between the colors of the tidal features
and their parent disks.   Lower limits are plotted as solid regions,
while upper limits are plotted as hatched regions.
}
\end{figure}

\begin{figure}
\plotone{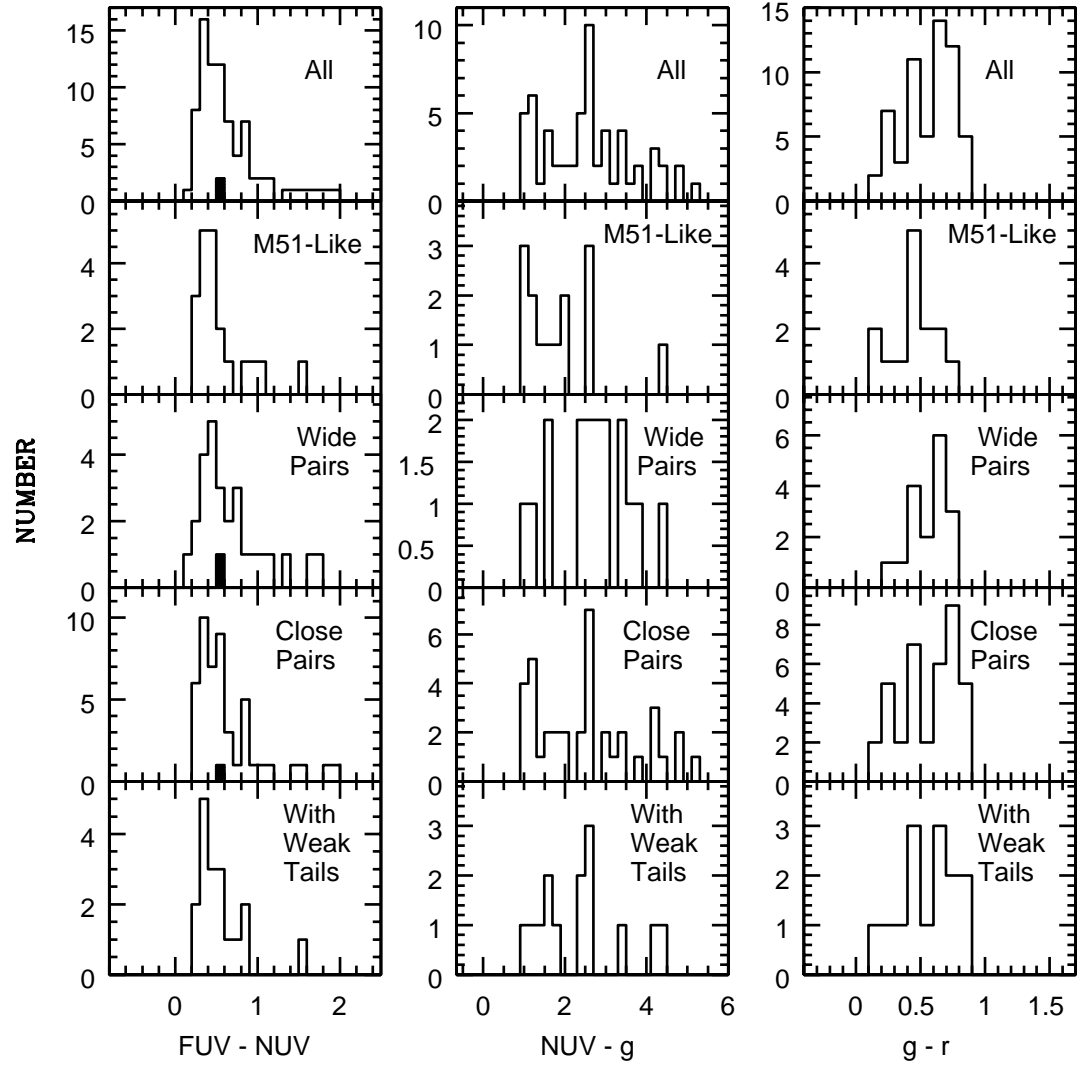}
\caption{
  \small 
Histograms of the FUV $-$ NUV, NUV $-$ g, and g $-$ r colors
of the disks of various subsets of the sample galaxies. 
The wide pairs are galaxies separated by $\ge$30 kpc.
The filled regions are lower limits.
}
\end{figure}

\begin{figure}
\plotone{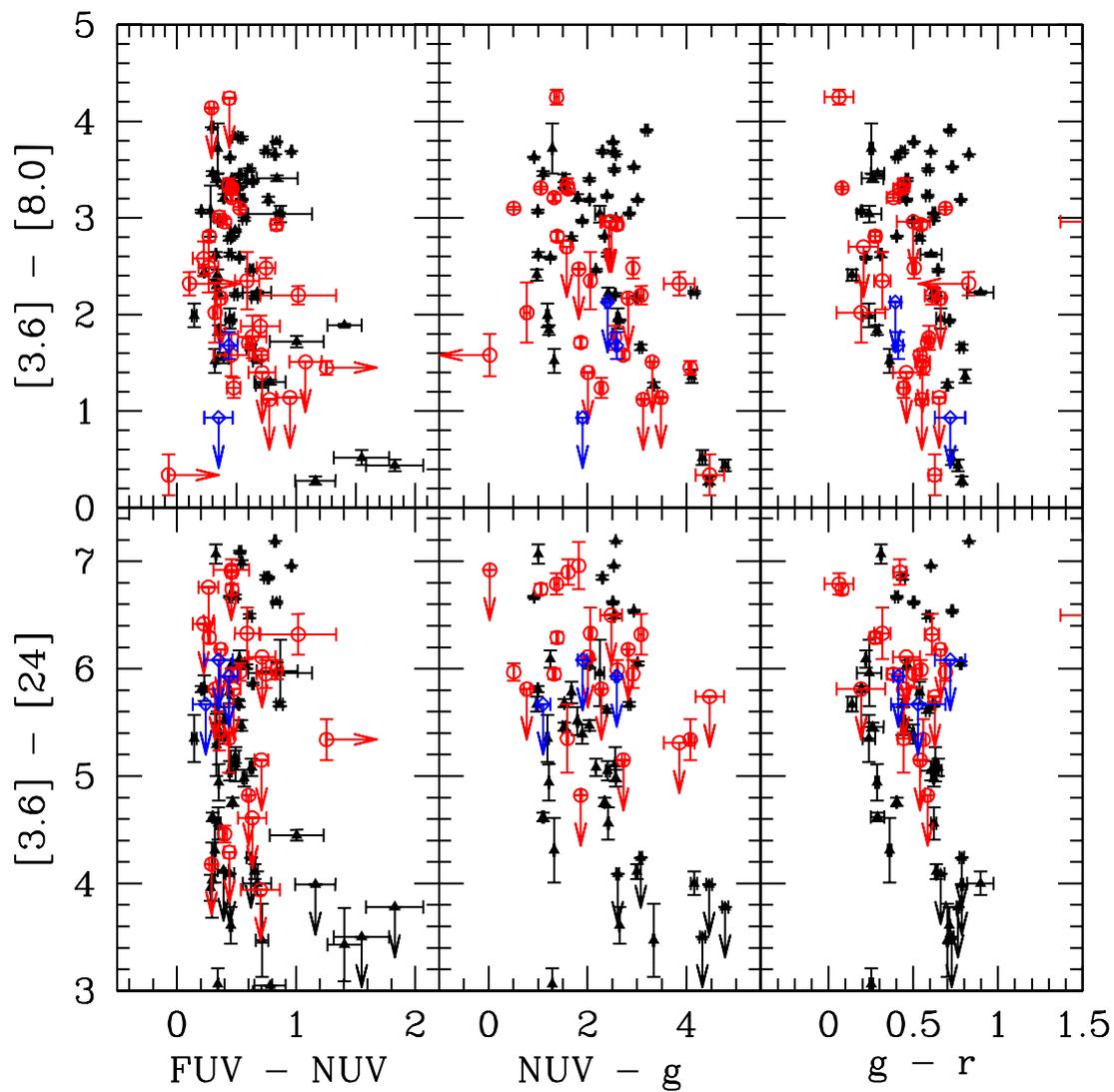}
\caption{
  \small 
Comparison of the UV/optical colors of the SB\&T disks (black filled
triangles), tidal features (red open circles), and candidate
TDGs (blue open diamonds) with the Spitzer
[3.6] $-$ [24] and [3.6] $-$ [8.0] colors.
}
\end{figure}

\begin{figure}
\plotone{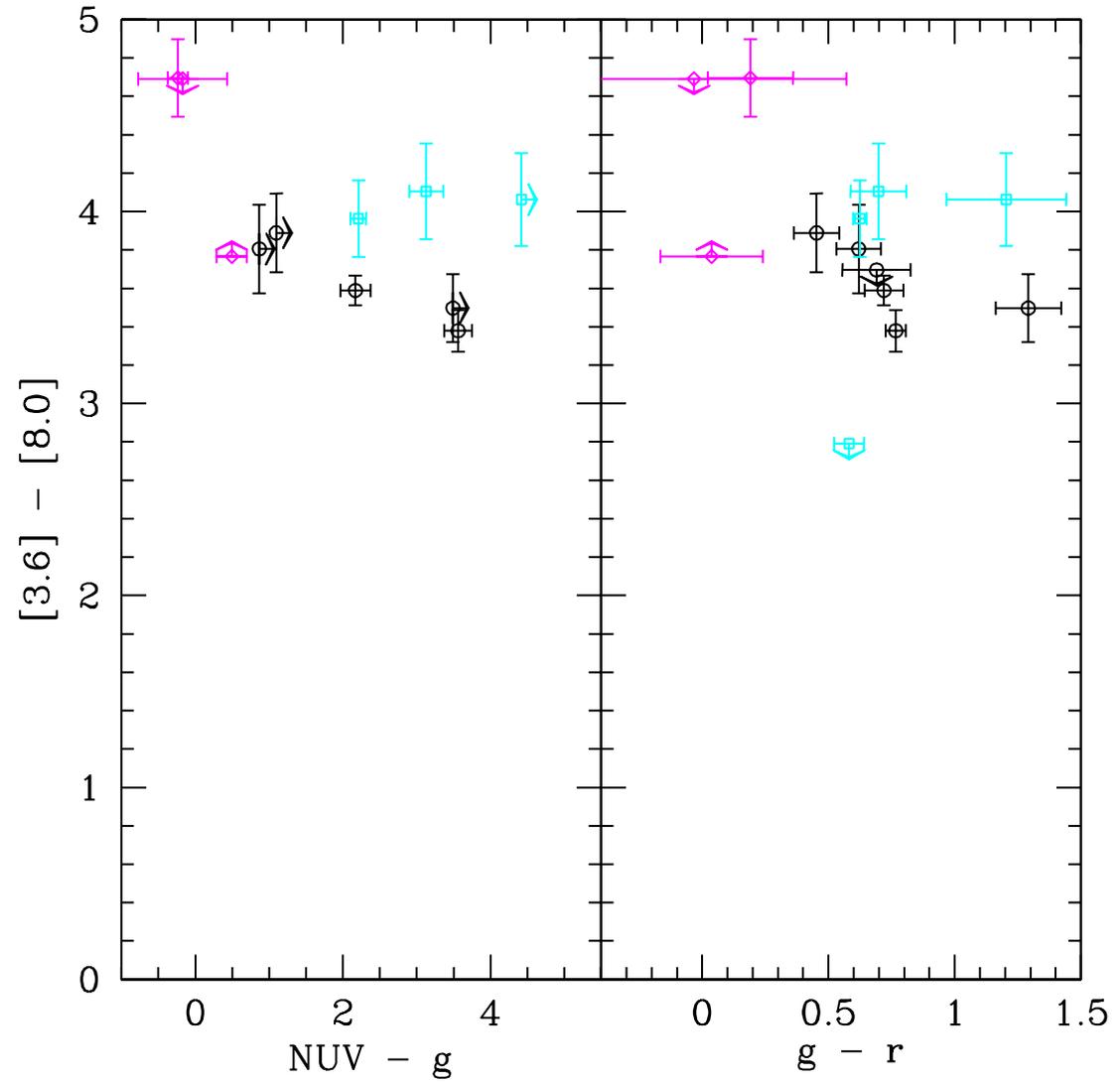}
\caption{
  \small 
For star forming clumps in Arp 285, 
comparison of the NUV $-$ g and g $-$ r colors with the 
Spitzer
[3.6] $-$ [8.0] colors.
The magenta open diamonds are points from the northern tail, the 
cyan open circles are clumps in the NGC 2856 disk, and 
the black open squares are clumps in the NGC 2854 disk.
Note that the axes for these plots are the same as the axes for
the two upper right panels of Figure 31.
}
\end{figure}

\begin{figure}
\plotone{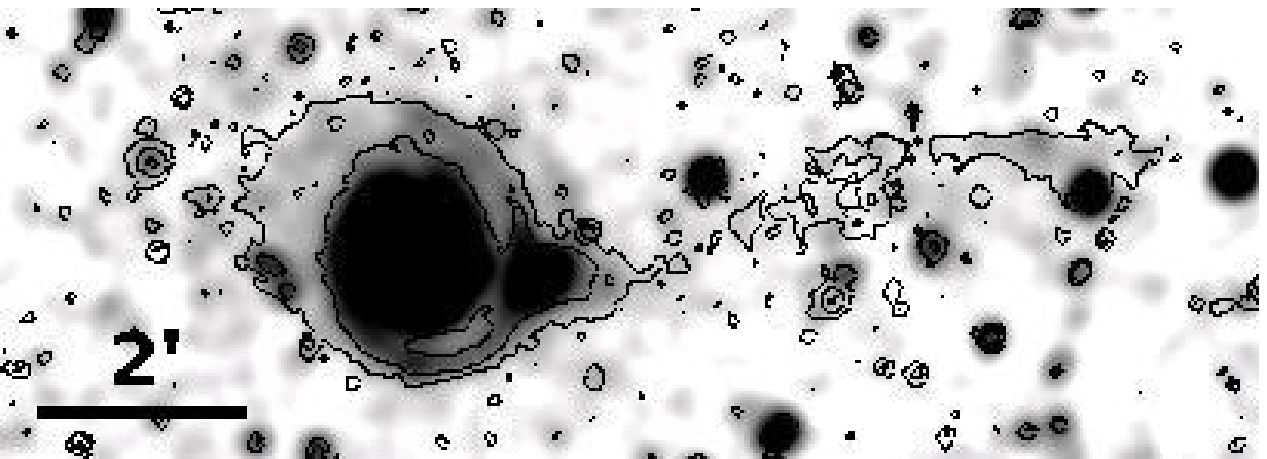}
\caption{
  \small 
Contours of the smoothed SDSS g image of Arp 297S, superimposed
on the smoothed GALEX NUV image.
}
\end{figure}

\clearpage
%
%
\LongTables
\begin{landscape}
\begin{deluxetable}{ccrrrccrrrrrrrrr}
\tabletypesize{\scriptsize}
\def\et#1#2#3{${#1}^{+#2}_{-#3}$}
\tablewidth{0pt}
\tablecaption{The Interacting Galaxy Sample\label{tab-1}}
\tablehead{
\multicolumn{1}{c}{System} &
\multicolumn{1}{c}{Other } &
\colhead{Distance} & 
\colhead{Separation} & 
\colhead{LOG L(FIR)$^a$} & 
\colhead{Notes on Morphology} &
\colhead{Nuclear Spectral} 
\\ 
\multicolumn{1}{c}{}
& \multicolumn{1}{c}{Names} 
& \multicolumn{1}{c}{(Mpc)} 
& \multicolumn{1}{c}{(kpc)} 
& \multicolumn{1}{c}{(L$_{\sun}$)} 
& \multicolumn{1}{c}{} 
& \multicolumn{1}{c}{Types} \\
}
\startdata
Arp 24 & NGC 3445 & 29 & 9.6 & 9.5 & M51-like, weak bridge & HII$^b$\\
Arp 34 & NGC 4613/4/5 & 67 & 43 & 10.1 & Equal-mass spirals, short tails, small third galaxy & $-$\\
Arp 35 & UGC 212  & 63 & 47 & 9.8 & M51-like & $-$  \\
Arp 65 & NGC 90/93 & 70 & 57 & 9.5 & Widely separated pair & $-$\\
Arp 72 & NGC 5994/6 & 47 & 21 & 10.2 & M51-like & LINER+HII$^b$\\
Arp 82 & NGC 2535/6 & 57& 28 & 10.2 & M51-like & HII/HII$^c$,LINER/HII$^b$\\
Arp 84 & NGC 5394/5 & 50 & 28 & 10.7 & M51-like, but bridge from smaller galaxy & HII/LINER$^{b,c}$\\
Arp 85 & M51,NGC 5194/5 & 6.2 & 8.0 & 9.7 & M51; spiral w/ small companion, bridge & LINER/HII$^{b,c}$\\
Arp 86 & NGC 7752/3 & 67 & 40 & 10.7 & M51-like & HII/LINER$^c$\\
Arp 87 & NGC 3808   & 97 & 29 & 10.8 & M51-like, but near-equal mass & LINER/HII$^{b,c}$\\
Arp 89 & NGC 2648   & 31 & 22 & 9 & M51-like & $-$\\
Arp 100 & IC 18/19 & 85 & 84 & $-$ & Wide equal mass pair, long tail & $-$ \\
Arp 101 & UGC 10164/9 & 64 & 44 & 9.7 &Wide equal mass pair, long tail & $-$ \\
Arp 105 & NGC 3561/UGC 06224 & 120 & 33 & 11 & Equal mass spiral/elliptical pair & $-$\\
Arp 107 & UGC 5984 & 143 & 46 & 10.1 & Ring-like spiral w/ small E, bridge,short tail & Seyfert$^c$\\
Arp 112 & NGC 7805/6 & 66 & 17 &9.3 & Unequal mass pair & $-$ \\
Arp 120 & NGC 4435/8 & 17 & 22 & 9.4 & $-$ & $-$ \\
Arp 173 & UGC 09561 & 127 & 31 & 11 & Close unequal mass pair, long tail & $-$ \\
Arp 181 & NGC 3212/5 & 129 & 46 & 10.6 & Two close equal mass spirals, long tail    & $-$\\
Arp 192 & NGC 3303 & 88 & 2 & 9.6 & Two close spirals, long tail & $-$ \\
Arp 202 & NGC 2719 & 44& 4.7 & 9.75 & Unequal mass pair & HII/HII$^c$\\
Arp 242 & NGC 4676 & 90 & 15 & 10.6 & Equal mass pair, two long tails & LINER/LINER$^c$, no emission$^b$ \\
Arp 244 & NGC 4038/9 & 27 & 8.6 & 10.8 & The Antennae; equal mass pair, long tails & HII/HII$^c$,HII/HII+LINER$^b$\\
Arp 245 & NGC 2992/3 & 38 & 32 & 10.5 & Equal mass separated spiral, prominent tails & $-$ \\
Arp 253 & UGC 173/4 & 29& 9.8 & 8.7 & Two close spirals, short tails & HII$^b$\\
Arp 254 & NGC 5917 & 28 & 34 & 9.7& Wide equal mass, long tail & $-$  \\
Arp 261 & $-$      & 28 & 7  & 9.3 & Close spiral pair & HII$^b$  \\
Arp 269 & NGC 4485/4490 & 8.6& 9  & 9.8 & M51-like & $-$  \\
Arp 270 & NGC 3395/6 & 27 & 9 & 9.9 & Two close spirals, short tail & HII/HII$^c$/Liner-HII/Liner-HII$^b$\\
Arp 271 & NGC 5426/7 & 40 & 20 & 10 & Close equal mass spirals; bridge, no tail & HII/Seyfert$^{b, c}$\\
Arp 280 & NGC 3769 & 12 & 4.6 & 8.9 & Unequal mass spirals, short tails & $-$\\
Arp 282 & NGC 169 & 60 & 7.1 & 10 & Close pair & no emission/HII+LINER$^c$\\
Arp 283 & NGC 2798/9 & 26 & 12 & 10.4 & Two close spirals, tails+bridge & HII/HII$^c$,HII/HII+LINER$^b$\\
Arp 284 & NGC 7714/5 & 34 & 20 & 10 & Unequal mass pair, partial ring, tails,bridge &  HII/HII$^{b, c}$\\
Arp 285 & NGC 2854/6 & 39 & 40 & 10 & Equal mass widely separated spirals & $-$\\
Arp 290 & IC 195/6 & 47& 30 & 9.2 & Unequal mass separated spirals & HII/LINER$^c$\\
Arp 295 & $-$     & 88 & 120 & 10.8 & Wide unequal mass pair, long bridge & LINER/HII$^c$\\
Arp 297N & NGC 5753/5 & 131 & 40 & 11 & Spiral with small companion & HII$^c$\\
Arp 297S & NGC 5752/4 & 64 & 35 & 10.2 & Spiral with small comanion & HII$^c$\\
Arp 298 & NGC 7469/IC5283 & 63 & 24 & 11.2 & Disk galaxies w/ ring, disturbed companion & Seyfert/HII$^c$\\
Arp 305 & NGC 4016/4017 & 47 & 80 & 9.7& Wide equal mass pair & $-$  \\
NGC 4567 & NGC 4567 & 36 & 12 & 10.7 & Two close spirals, no tails & $-$\\
\enddata
\tablenotetext{a}{Includes the IRAS flux from both galaxies in the pair.}
\tablenotetext{b}{Nuclear spectral type from \citet{dahari85}.}
\tablenotetext{c}{Nuclear spectral type from \citet{keel85}.}

\end{deluxetable}
\end{landscape}
%

\clearpage




\begin{table}
\begin{scriptsize}
\caption{\bf{GALEX Observations}}
\begin{center}
\begin{tabular}{ccccrrc}
\hline
System & Min. & Max. & Tile & NUV & FUV & SDSS? \\
Name\footnotemark[1] & Obs. Date & Obs. Date & Name & (sec) & (sec) & \\
\hline
Arp 24 & 1/31/2004 & 5/2/2004 & LOCK\_O5 & 85021 & 28080 & yes\\
Arp 34 & 5/23/2007 & 4/3/2008 & GI1\_026017\_Arp34 & 2773 & 2773 & yes \\
Arp 35 & 9/8/2003 & 10/30/2007 & MISDR1\_29132\_0390 & 3354 & 3354&  \\
Arp 65 & 10/13/2004 & 11/2/2005 & GI1\_026002\_Arp65 & 5707 & 3310& \\
Arp 72 &  5/20/2005 & 6/4/2007 & GI1\_026024\_Arp72 & 4899 & 2609& yes \\
Arp 82 & 2/13/2005 & 2/25/2005 & GI1\_026007\_Arp82 & 3012 & 1680& yes \\
Arp 84 & 4/12/2006 & 4/2/2008 & GI1\_026018\_Arp84 & 4286 & 2820 & yes \\
Arp 85  & 5/29/2007 & 5/9/2008 & 
GI3\_050006\_NGC5194 NGA\_M51 & 10136 & 10136 & yes \\
Arp 86 & 10/2/2004 & 8/13/2005 & GI1\_026027\_Arp86 & 3221 & 1679 &  \\
Arp 87  & 3/26/2006 & 3/26/2006 & GI1\_026015\_Arp87 & 1595 & 1595 & yes \\
Arp 89  & 2/18/2005 & 2/18/2005 & GI1\_026008\_Arp89 & 1685 & 1685 & yes \\
Arp 100 & 10/1/2004 & 10/1/2004 & GI1\_026004\_Arp100 & 1600 & 1600&  \\
Arp 101 & 6/17/2005 & 5/21/2007 & GI1\_026025\_Arp101 & 5054 & 2326 & yes \\
Arp 105 & 4/8/2005 & 4/8/2005 & NGA\_Arp105 & 959 & $-$ & yes \\
Arp 107 & 4/7/2005 & 3/25/2006 & GI1\_026013\_Arp107 & 2610 & 1094 &yes \\
Arp 112 & 9/25/2004 & 9/25/2004 & GI1\_026001\_Arp112 & 1648 & 1648 & yes\\
Arp 120 & 3/12/2004 & 4/28/2005 & NGA\_Virgo\_MOS09 & 4575 & 1422 & yes \\
Arp 173 & 5/16/2005 & 6/1/2007 & GI1\_037001\_J144938p091058 & 11717 &	5111 & yes \\
Arp 181 & 3/7/2005 & 3/7/2005 & GI1\_026011\_Arp181 & 1599 & 1599& yes \\
Arp 192 & 2/6/2006 & 2/6/2006 & GI1\_026012\_Arp192 & 1561 & 1561 & yes\\
Arp 202 & 1/21/2006 & 1/28/2006 & GI1\_026009\_Arp202 & 2693 & 2693 & yes \\
Arp 242 & 4/8/2008 & 4/8/2008 & GI1\_077006\_BDp312402 & 1669 & 1669 & yes \\
Arp 244 & 2/22/2004 & 2/28/2004 & NGA\_Antennae & 2365 & 2365 & \\
Arp 245 & 1/7/2004 & 2/21/2005 & NGA\_Arp245  & 4283 &	1048 & \\
Arp 253 & 3/16/2005 & 2/17/2006 & GI1\_026010\_Arp253 & 4709 & 3004 & \\
Arp 254 & 5/5/2006 & 6/3/2007 & GI1\_026023\_Arp254 & 4583 & 2919 & \\
Arp 261 & 5/4/2006 & 6/2/2007 & GI1\_026021\_Arp261 & 3844 & 2157 & \\
Arp 269 & 3/26/2005 & 3/20/2008 & NGA\_NGC4490 & 4471 & 3247 & yes \\
Arp 270 & 3/23/2006 & 3/12/2007 & GI1\_078004\_NGC3395 & 2660 &	1487 & yes \\
Arp 271 & 5/4/2006 & 5/23/2006 & GI1\_009087\_NGC5426 & 3001 & 1312 &  \\
Arp 280 & 2/26/2006 & 2/26/2006 & GI1\_026014\_Arp280 & 1660 & 1660 & yes \\
Arp 282 & 11/7/2004 & 11/4/2005 & GI1\_026005\_Arp282  & 3518 & 1996 & yes \\
Arp 283 & 2/8/2004 & 2/4/2007 & NGA\_NGC2798 & 4272 & 2798 & yes \\
Arp 284 & 10/12/2004 & 9/13/2006 & GI1\_045006\_Arp284 & --- & 4736 & \\
Arp 285 & 4/21/2005 & 4/21/2005 & MISDR3\_03371\_0900	& 1419 & --- & yes \\
Arp 290 & 10/15/2003 & 10/15/2003 & MISDR1\_17381\_0427 & 1669 & 1669 & yes \\
Arp 295 & 10/5/2003 & 10/5/2003 & NGRG\_A295 & 1233 & 1233 & \\
Arp 297 & 5/9/2006 & 5/17/2007 & GI1\_026020\_Arp297 & 4639 & 2962 & yes\\
Arp 298 & 9/4/2003 & 10/3/2003 & NGA\_NGC7469 & 3768 & 3768 & \\
Arp 305 & 3/27/2006 & 3/27/2006 & GI1\_026016\_Arp305 & 1416 & 1416 & yes \\
NGC 4567 & 4/5/2005 & 4/29/2005 & NGA\_Virgo\_MOS07 & 2928 & --- & yes\\
\hline
\end{tabular}
\end{center}
\end{scriptsize}
\end{table}

%
%
\LongTables
\begin{landscape}
\begin{deluxetable}{cccccccccc}
\tabletypesize{\scriptsize}
\setlength{\tabcolsep}{0.03in}
\def\et#1#2#3{${#1}^{+#2}_{-#3}$}
\tablewidth{0pt}
\tablecaption{Optical and UV Magnitudes for Interacting Galaxy Sample\label{tab-4}}
\tablehead{
\multicolumn{1}{c}{} &
\multicolumn{1}{c}{} &
\multicolumn{1}{c}{} &
\colhead{FUV} & 
\colhead{NUV} & 
\colhead{u} & 
\colhead{g} & 
\colhead{r} & 
\colhead{i} & 
\colhead{z} 
\\ 
\multicolumn{1}{c}{Arp Name} &
\multicolumn{1}{c}{Component} &
\multicolumn{1}{c}{Other Name} &
\colhead{(mag)} & 
\colhead{(mag)} & 
\colhead{(mag)} & 
\colhead{(mag)} & 
\colhead{(mag)} & 
\colhead{(mag)} & 
\colhead{(mag)} 
\\ 
}
\startdata
  Arp 24 &        E &      UGC     6021 &     17.89 $\pm$    0.01 &     17.57 $\pm$    0.00 &     17.24 $\pm$    0.03 &     16.25 $\pm$    0.01 &     15.88 $\pm$    0.01 &     15.75 $\pm$    0.02 &     15.80 $\pm$    0.06   \\
  Arp 24 &        MAIN &      NGC     3445 &     13.89 $\pm$    0.00 &     13.69 $\pm$    0.00 &     13.41 $\pm$    0.00 &     12.69 $\pm$    0.00 &     12.49 $\pm$    0.00 &     12.40 $\pm$    0.01 &     12.38 $\pm$    0.02   \\
  Arp 34 &       NE &      NGC     4615 &     16.41 $\pm$    0.00 &     15.90 $\pm$    0.00 &     15.32 $\pm$    0.03 &     14.37 $\pm$    0.00 &     13.93 $\pm$    0.00 &     13.68 $\pm$    0.00 &     13.49 $\pm$    0.01   \\
  Arp 34 &    NE E TAIL &      NGC 4615 EAST TAIL &     16.93 $\pm$    0.01 &     16.46 $\pm$    0.00 &     16.28 $\pm$    0.02 &     15.41 $\pm$    0.00 &     15.33 $\pm$    0.00 &     15.12 $\pm$    0.01 &     14.87 $\pm$    0.02   \\
  Arp 34 &    NE W TAIL &      NGC 4615 WEST TAIL &     16.97 $\pm$    0.01 &     16.44 $\pm$    0.00 &     16.64 $\pm$    0.02 &     15.93 $\pm$    0.00 &     15.24 $\pm$    0.00 &     15.09 $\pm$    0.00 &     15.10 $\pm$    0.02   \\
  Arp 34 & NW SMALL &      NGC     4613 &     18.02 $\pm$    0.02 &     17.39 $\pm$    0.00 &     16.57 $\pm$    0.05 &     15.22 $\pm$    0.00 &     14.57 $\pm$    0.00 &     14.26 $\pm$    0.00 &     14.01 $\pm$    0.01   \\
  Arp 34 &        S &      NGC     4614 &     17.89 $\pm$    0.05 &     17.17 $\pm$    0.01 &     15.52 $\pm$    0.09 &     13.84 $\pm$    0.00 &     13.13 $\pm$    0.00 &     12.77 $\pm$    0.01 &     12.51 $\pm$    0.01   \\
  Arp 35 &         BRIDGE &      UGC     212 BRIDGE &     18.01 $\pm$    0.01 &     17.83 $\pm$    0.01 &        $-$ &        $-$ &        $-$ &        $-$ &        $-$   \\
  Arp 35 &        N &      UGC      212 &     15.90 $\pm$    0.00 &     15.61 $\pm$    0.00 &        $-$ &        $-$ &        $-$ &        $-$ &        $-$   \\
  Arp 35 &       N TAIL &      UGC 212 NORTH TAIL &     18.13 $\pm$    0.02 &     17.95 $\pm$    0.02 &        $-$ &        $-$ &        $-$ &        $-$ &        $-$   \\
  Arp 35 &        S &      UGC 212notes01 &     18.04 $\pm$    0.01 &     17.73 $\pm$    0.01 &        $-$ &        $-$ &        $-$ &        $-$ &        $-$   \\
  Arp 35 &       S TAIL &      UGC 212notes01 SOUTH TAIL &     21.26 $\pm$    0.04 &     21.16 $\pm$    0.05 &        $-$ &        $-$ &        $-$ &        $-$ &        $-$   \\
  Arp 65 &        E &      NGC       93 &     18.69 $\pm$    0.13 &     17.91 $\pm$    0.04 &        $-$ &        $-$ &        $-$ &        $-$ &        $-$   \\
  Arp 65 &       N TAIL &      NGC      90 TAIL &     19.24 $\pm$    0.07 &     18.97 $\pm$    0.04 &        $-$ &        $-$ &        $-$ &        $-$ &        $-$   \\
  Arp 65 &       S TAIL &      NGC      90 TAIL &     18.77 $\pm$    0.08 &     18.54 $\pm$    0.04 &        $-$ &        $-$ &        $-$ &        $-$ &        $-$   \\
  Arp 65 &        W &      NGC       90 &     17.59 $\pm$    0.04 &     17.20 $\pm$    0.02 &        $-$ &        $-$ &        $-$ &        $-$ &        $-$   \\
  Arp 72 &         BRIDGE &      NGC  5994/6 BRIDGE &     16.54 $\pm$    0.01 &     16.08 $\pm$    0.01 &     15.96 $\pm$    0.08 &     14.76 $\pm$    0.03 &     14.38 $\pm$    0.03 &     14.21 $\pm$    0.04 &     14.33 $\pm$    0.08   \\
  Arp 72 &        E &      NGC     5996 &     14.40 $\pm$    0.00 &     13.95 $\pm$    0.00 &     13.82 $\pm$    0.01 &     13.03 $\pm$    0.01 &     12.63 $\pm$    0.01 &     12.54 $\pm$    0.01 &     12.51 $\pm$    0.02   \\
  Arp 72 &       E TAIL &      NGC    5996 TAIL &     15.94 $\pm$    0.01 &     15.67 $\pm$    0.00 &     15.37 $\pm$    0.05 &     14.28 $\pm$    0.02 &     14.01 $\pm$    0.02 &     13.95 $\pm$    0.04 &     14.07 $\pm$    0.06   \\
  Arp 72 &        W &      NGC     5994 &     16.43 $\pm$    0.00 &     16.10 $\pm$    0.00 &     16.01 $\pm$    0.04 &     15.12 $\pm$    0.01 &     14.98 $\pm$    0.02 &     14.89 $\pm$    0.03 &     14.92 $\pm$    0.05   \\
  Arp 82 &         BRIDGE &      NGC  2535/6 BRIDGE &     17.33 $\pm$    0.01 &     16.90 $\pm$    0.01 &     16.34 $\pm$    0.08 &     15.25 $\pm$    0.02 &     14.82 $\pm$    0.03 &     14.66 $\pm$    0.05 &     14.77 $\pm$    0.05   \\
Arp 82 & E ARC & NGC 2535/6 EAST ARC & 17.98 $\pm$ 0.03 & 17.63 $\pm$ 0.02 & 16.54 $\pm$ 0.16 & 14.75 $\pm$ 0.02 & 14.08 $\pm$ 0.02 & 13.94 $\pm$ 0.04 & 13.60 $\pm$ 0.03 \\
  Arp 82 &        N &      NGC     2535 &     15.17 $\pm$    0.00 &     14.71 $\pm$    0.00 &     13.76 $\pm$    0.02 &     12.65 $\pm$    0.00 &     12.20 $\pm$    0.01 &     11.95 $\pm$    0.01 &     11.81 $\pm$    0.01   \\
  Arp 82 &       N TAIL &      NGC 2535 NORTH TAIL &     17.50 $\pm$    0.05 &     17.05 $\pm$    0.03 &     $\ge$16.09     &     15.89 $\pm$    0.13 &     15.28 $\pm$    0.19 &     15.05 $\pm$    0.33 &     $\ge$15.19       \\
  Arp 82 &        S &      NGC     2536 &     16.49 $\pm$    0.01 &     16.10 $\pm$    0.00 &     15.39 $\pm$    0.05 &     14.30 $\pm$    0.01 &     13.87 $\pm$    0.02 &     13.65 $\pm$    0.03 &     13.60 $\pm$    0.02   \\
  Arp 84 &         BRIDGE &      NGC  5394/5 BRIDGE &     18.84 $\pm$    0.04 &     18.00 $\pm$    0.01 &     16.94 $\pm$    0.04 &     15.40 $\pm$    0.01 &     14.86 $\pm$    0.01 &     14.58 $\pm$    0.01 &     14.66 $\pm$    0.06   \\
  Arp 84 &        N &      NGC     5394 &     17.37 $\pm$    0.01 &     16.40 $\pm$    0.00 &     15.11 $\pm$    0.01 &     13.88 $\pm$    0.00 &     13.27 $\pm$    0.00 &     12.96 $\pm$    0.00 &     12.70 $\pm$    0.01   \\
  Arp 84 &       N TAIL &      NGC    5394 TAIL &     18.76 $\pm$    0.08 &     18.02 $\pm$    0.01 &     16.89 $\pm$    0.06 &     15.10 $\pm$    0.01 &     14.59 $\pm$    0.01 &     14.37 $\pm$    0.01 &     14.49 $\pm$    0.10   \\
  Arp 84 &        S &      NGC     5395 &     15.15 $\pm$    0.02 &     14.59 $\pm$    0.00 &     13.46 $\pm$    0.01 &     12.01 $\pm$    0.00 &     11.39 $\pm$    0.00 &     11.05 $\pm$    0.00 &     10.88 $\pm$    0.02   \\
  Arp 84 &      S N TAIL &      NGC   5395 N TAIL &     20.27 $\pm$    0.10 &     19.68 $\pm$    0.03 &     $\ge$19.26     &     17.62 $\pm$    0.03 &     17.30 $\pm$    0.04 &     16.90 $\pm$    0.05 &     $\ge$16.81       \\
  Arp 85 &         BRIDGE &      NGC  5194/5 BRIDGE &     14.78 $\pm$    0.02 &     14.38 $\pm$    0.00 &        $-$ &        $-$ &        $-$ &        $-$ &        $-$   \\
  Arp 85 &        N &      NGC     5195 &     15.71 $\pm$    0.23 &     14.70 $\pm$    0.02 &        $-$ &        $-$ &        $-$ &        $-$ &        $-$   \\
  Arp 85 &        S &      NGC     5194 &     11.55 $\pm$    0.01 &     10.92 $\pm$    0.00 &        $-$ &        $-$ &        $-$ &        $-$ &        $-$   \\
  Arp 86 &         BRIDGE &      NGC  7752/3 BRIDGE &     17.77 $\pm$    0.04 &     17.41 $\pm$    0.02 &        $-$ &        $-$ &        $-$ &        $-$ &        $-$   \\
  Arp 86 &        N &      NGC     7753 &     15.56 $\pm$    0.03 &     15.07 $\pm$    0.01 &        $-$ &        $-$ &        $-$ &        $-$ &        $-$   \\
  Arp 86 &        S &      NGC     7752 &     16.81 $\pm$    0.02 &     16.32 $\pm$    0.01 &        $-$ &        $-$ &        $-$ &        $-$ &        $-$   \\
  Arp 87 &         BRIDGE &      NGC    3808 BRIDGE &     18.96 $\pm$    0.03 &     18.30 $\pm$    0.01 &     17.93 $\pm$    0.08 &     16.45 $\pm$    0.01 &     16.21 $\pm$    0.01 &     15.92 $\pm$    0.01 &     15.90 $\pm$    0.05   \\
  Arp 87 &        N &      NGC   3808B &     18.08 $\pm$    0.03 &     17.25 $\pm$    0.01 &     16.09 $\pm$    0.03 &     14.74 $\pm$    0.01 &     14.23 $\pm$    0.00 &     13.93 $\pm$    0.00 &     13.80 $\pm$    0.01   \\
  Arp 87 &        S &      NGC    3808A &     16.63 $\pm$    0.01 &     16.09 $\pm$    0.00 &     15.28 $\pm$    0.02 &     14.05 $\pm$    0.00 &     13.59 $\pm$    0.00 &     13.34 $\pm$    0.00 &     13.19 $\pm$    0.01   \\
  Arp 87 &  SMALL N &                &     22.00 $\pm$    0.26 &     21.13 $\pm$    0.07 &     $\ge$19.95     &     18.89 $\pm$    0.05 &     18.65 $\pm$    0.05 &     18.12 $\pm$    0.05 &     $\ge$18.40       \\
  Arp 87 &       S TAIL &      NGC   3808A TAIL &     20.40 $\pm$    0.14 &     19.37 $\pm$    0.03 &     $\ge$19.67     &     18.21 $\pm$    0.04 &     17.65 $\pm$    0.02 &     17.40 $\pm$    0.04 &     17.94 $\pm$    0.23   \\
  Arp 89 &        E &      KPG      168 &     17.25 $\pm$    0.01 &     16.82 $\pm$    0.01 &     16.29 $\pm$    0.06 &     15.14 $\pm$    0.01 &     14.60 $\pm$    0.01 &     14.34 $\pm$    0.02 &     14.25 $\pm$    0.03   \\
  Arp 89 &       E TAIL &      KPG 168 EAST TAIL &     20.10 $\pm$    0.10 &     19.38 $\pm$    0.05 &     $\ge$18.46     &     17.37 $\pm$    0.06 &     16.92 $\pm$    0.05 &     16.57 $\pm$    0.07 &     16.65 $\pm$    0.17   \\
  Arp 89 &        W &      NGC     2648 &     18.08 $\pm$    0.23 &     16.53 $\pm$    0.06 &     14.16 $\pm$    0.08 &     12.21 $\pm$    0.01 &     11.48 $\pm$    0.01 &     11.06 $\pm$    0.01 &     10.86 $\pm$    0.01   \\
  Arp 100 &        N &       IC       18 &     19.69 $\pm$    0.10 &     18.77 $\pm$    0.03 &        $-$ &        $-$ &        $-$ &        $-$ &        $-$   \\
  Arp 100 &       N TAIL &       IC 18 NORTH TAIL &     19.34 $\pm$    0.27 &     18.83 $\pm$    0.11 &        $-$ &        $-$ &        $-$ &        $-$ &        $-$   \\
  Arp 100 &        S &       IC       19 &     21.12 $\pm$    0.25 &     19.33 $\pm$    0.04 &        $-$ &        $-$ &        $-$ &        $-$ &        $-$   \\
  Arp 100 &       S TAIL &       IC 18 SOUTH TAIL &     20.03 $\pm$    0.29 &     19.71 $\pm$    0.14 &        $-$ &        $-$ &        $-$ &        $-$ &        $-$   \\
  Arp 101 &         BRIDGE &      UGC 10164/9 BRIDGE &     $\ge$18.45     &     18.92 $\pm$    0.15 &     17.87 $\pm$    0.33 &     15.33 $\pm$    0.03 &     14.93 $\pm$    0.03 &     14.44 $\pm$    0.02 &     14.83 $\pm$    0.05   \\
  Arp 101 &        N &      UGC    10169 &     18.73 $\pm$    0.28 &     17.41 $\pm$    0.03 &     15.49 $\pm$    0.03 &     13.89 $\pm$    0.01 &     13.45 $\pm$    0.01 &     13.12 $\pm$    0.01 &     12.87 $\pm$    0.01   \\
  Arp 101 &       N TAIL &      UGC 10169 NORTH TAIL &     $\ge$17.85     &     18.81 $\pm$    0.22 &     $\ge$17.38     &     16.07 $\pm$    0.10 &     15.45 $\pm$    0.08 &     15.17 $\pm$    0.07 &     15.41 $\pm$    0.15   \\
  Arp 101 &        S &      UGC    10164 &     $\ge$18.44     &     17.90 $\pm$    0.05 &     15.74 $\pm$    0.15 &     14.13 $\pm$    0.02 &     13.57 $\pm$    0.02 &     13.27 $\pm$    0.02 &     13.18 $\pm$    0.02   \\
  Arp 105 &       E TAIL &       VV     237d &        $-$ &     19.26 $\pm$    0.05 &     $\ge$18.93     &     17.68 $\pm$    0.05 &     17.47 $\pm$    0.07 &     17.03 $\pm$    0.08 &     $\ge$17.02       \\
  Arp 105 &        N &      NGC    3561A &        $-$ &     18.04 $\pm$    0.03 &     16.49 $\pm$    0.05 &     14.85 $\pm$    0.01 &     14.13 $\pm$    0.01 &     13.73 $\pm$    0.01 &     13.56 $\pm$    0.02   \\
  Arp 105 &       N TAIL &      NGC 3561A NORTH TAIL &        $-$ &     18.55 $\pm$    0.13 &     $\ge$17.20     &     16.11 $\pm$    0.07 &     15.61 $\pm$    0.07 &     15.28 $\pm$    0.09 &     $\ge$15.16       \\
  Arp 105 &        S &      NGC     3561 &        $-$ &     18.45 $\pm$    0.06 &     16.30 $\pm$    0.05 &     14.35 $\pm$    0.01 &     13.54 $\pm$    0.00 &     13.13 $\pm$    0.01 &     12.94 $\pm$    0.02   \\
  Arp 105 &       S TAIL &      NGC 3561 SOUTH TAIL &        $-$ &     18.52 $\pm$    0.03 &     17.59 $\pm$    0.09 &     15.98 $\pm$    0.01 &     15.38 $\pm$    0.01 &     15.01 $\pm$    0.01 &     15.10 $\pm$    0.05   \\
  Arp 105 &         TDG &                   &        $-$ &     18.82 $\pm$    0.04 &     18.33 $\pm$    0.20 &     16.42 $\pm$    0.02 &     16.02 $\pm$    0.02 &     15.83 $\pm$    0.04 &     16.35 $\pm$    0.19   \\
  Arp 107 &         BRIDGE &      UGC    5984 BRIDGE &     20.26 $\pm$    0.05 &     19.55 $\pm$    0.03 &     $\ge$19.03     &     16.83 $\pm$    0.01 &     16.28 $\pm$    0.01 &     15.92 $\pm$    0.03 &     15.89 $\pm$    0.14   \\
  Arp 107 &        N &      UGC    5984N &     20.61 $\pm$    0.16 &     19.45 $\pm$    0.06 &     17.14 $\pm$    0.12 &     14.99 $\pm$    0.01 &     14.21 $\pm$    0.00 &     13.81 $\pm$    0.01 &     13.60 $\pm$    0.05   \\
  Arp 107 &       N TAIL &      UGC   5984N TAIL &     $\ge$21.63     &     21.70 $\pm$    0.28 &     $\ge$18.55     &     17.23 $\pm$    0.03 &     16.60 $\pm$    0.02 &     16.14 $\pm$    0.06 &     16.19 $\pm$    0.31   \\
  Arp 107 &        S &      UGC    5984S &     17.09 $\pm$    0.01 &     16.60 $\pm$    0.01 &     15.91 $\pm$    0.08 &     14.04 $\pm$    0.01 &     13.41 $\pm$    0.00 &     13.07 $\pm$    0.01 &     12.91 $\pm$    0.06   \\
  Arp 107 &       W TAIL &      UGC   5984S TAIL &     20.13 $\pm$    0.05 &     19.66 $\pm$    0.03 &     $\ge$19.06     &     17.38 $\pm$    0.02 &     16.93 $\pm$    0.02 &     16.71 $\pm$    0.06 &     16.69 $\pm$    0.29   \\
  Arp 112 &        N &      NGC     7806 &     17.72 $\pm$    0.07 &     17.27 $\pm$    0.01 &     15.52 $\pm$    0.05 &     13.90 $\pm$    0.00 &     13.18 $\pm$    0.00 &     12.77 $\pm$    0.01 &     12.51 $\pm$    0.00   \\
  Arp 112 &       N TAIL &      NGC 7806 NORTH TAIL &     $\ge$19.61     &     19.61 $\pm$    0.06 &     $\ge$17.93     &     16.80 $\pm$    0.02 &     16.12 $\pm$    0.03 &     15.74 $\pm$    0.05 &     15.78 $\pm$    0.05   \\
  Arp 112 &        S &      NGC     7805 &     $\ge$19.76     &     19.21 $\pm$    0.03 &     15.98 $\pm$    0.04 &     14.09 $\pm$    0.00 &     13.28 $\pm$    0.00 &     12.85 $\pm$    0.00 &     12.55 $\pm$    0.00   \\
  Arp 112 &         TDG &      KUG 2359+311 &     19.62 $\pm$    0.29 &     18.98 $\pm$    0.03 &     17.96 $\pm$    0.29 &     16.20 $\pm$    0.01 &     15.66 $\pm$    0.02 &     15.34 $\pm$    0.04 &     15.29 $\pm$    0.03   \\
  Arp 120 &         BRIDGE &      NGC  4435/8 BRIDGE &     18.70 $\pm$    0.12 &     17.75 $\pm$    0.02 &     16.48 $\pm$    0.23 &     14.26 $\pm$    0.03 &     13.61 $\pm$    0.02 &     13.27 $\pm$    0.02 &     13.17 $\pm$    0.05   \\
  Arp 120 &        N &      NGC     4435 &     17.96 $\pm$    0.14 &     16.02 $\pm$    0.01 &     13.09 $\pm$    0.02 &     11.30 $\pm$    0.00 &     10.54 $\pm$    0.00 &     10.13 $\pm$    0.00 &      9.92 $\pm$    0.01   \\
  Arp 120 &       N TAIL &      NGC 4438 NORTH TAIL &     17.00 $\pm$    0.07 &     16.23 $\pm$    0.02 &     14.99 $\pm$    0.16 &     13.11 $\pm$    0.03 &     12.56 $\pm$    0.02 &     12.21 $\pm$    0.02 &     12.08 $\pm$    0.05   \\
  Arp 120 &        S &      NGC     4438 &     15.51 $\pm$    0.02 &     14.68 $\pm$    0.00 &     12.77 $\pm$    0.03 &     10.97 $\pm$    0.00 &     10.20 $\pm$    0.00 &      9.76 $\pm$    0.00 &      9.55 $\pm$    0.01   \\
  Arp 120 &       S TAIL &      NGC 4438 SOUTH TAIL &     18.82 $\pm$    0.13 &     17.74 $\pm$    0.02 &     16.37 $\pm$    0.21 &     14.43 $\pm$    0.03 &     13.88 $\pm$    0.03 &     13.58 $\pm$    0.03 &     13.48 $\pm$    0.06   \\
  Arp 120 &       W TAIL &      NGC 4438 WEST TAIL &     19.00 $\pm$    0.18 &     19.03 $\pm$    0.09 &     $\ge$16.50     &     $\ge$16.53     &     $\ge$16.22     &     $\ge$15.99     &     $\ge$14.94       \\
  Arp 173 &         BRIDGE &      UGC     9561 &     21.54 $\pm$    0.25 &     20.64 $\pm$    0.08 &     $\ge$18.90     &     18.10 $\pm$    0.06 &     17.46 $\pm$    0.08 &     16.86 $\pm$    0.07 &     $\ge$17.00       \\
  Arp 173 &        N &      UGC 9561notes01 &     19.54 $\pm$    0.11 &     17.91 $\pm$    0.02 &     16.22 $\pm$    0.05 &     14.46 $\pm$    0.00 &     13.83 $\pm$    0.01 &     13.44 $\pm$    0.01 &     13.22 $\pm$    0.02   \\
  Arp 173 &        S &      UGC     9561 &     19.60 $\pm$    0.04 &     18.52 $\pm$    0.01 &     17.36 $\pm$    0.07 &     15.67 $\pm$    0.01 &     15.06 $\pm$    0.01 &     14.70 $\pm$    0.01 &     14.51 $\pm$    0.03   \\
  Arp 173 &       S TAIL &      UGC     9561 &     21.49 $\pm$    0.29 &     19.95 $\pm$    0.05 &     $\ge$18.64     &     17.44 $\pm$    0.04 &     16.95 $\pm$    0.06 &     16.59 $\pm$    0.07 &     $\ge$16.78       \\
  Arp 181 &        E &      NGC     3215 &     17.24 $\pm$    0.02 &     16.59 $\pm$    0.02 &     15.45 $\pm$    0.06 &     13.60 $\pm$    0.01 &     12.96 $\pm$    0.00 &     12.60 $\pm$    0.01 &     12.45 $\pm$    0.02   \\
  Arp 181 &        FAR WEST &                   &     18.47 $\pm$    0.01 &     18.32 $\pm$    0.02 &     18.36 $\pm$    0.16 &     17.13 $\pm$    0.02 &     16.89 $\pm$    0.03 &     16.76 $\pm$    0.05 &     17.53 $\pm$    0.32   \\
  Arp 181 &        W &      NGC     3212 &     17.85 $\pm$    0.02 &     16.99 $\pm$    0.02 &     15.95 $\pm$    0.07 &     14.14 $\pm$    0.01 &     13.52 $\pm$    0.01 &     13.15 $\pm$    0.01 &     12.98 $\pm$    0.02   \\
  Arp 181 &       W TAIL &      NGC    3212 TAIL &     19.81 $\pm$    0.10 &     18.84 $\pm$    0.09 &     16.78 $\pm$    0.12 &     14.58 $\pm$    0.01 &     14.26 $\pm$    0.01 &     13.97 $\pm$    0.01 &     13.87 $\pm$    0.04   \\
  Arp 181 &         TDG &                   &     20.73 $\pm$    0.07 &     20.49 $\pm$    0.08 &     $\ge$19.40     &     19.39 $\pm$    0.12 &     18.86 $\pm$    0.11 &     18.91 $\pm$    0.23 &     $\ge$17.82       \\
  Arp 192 &        N &      NGC 3303NED01 &     20.04 $\pm$    0.12 &     18.89 $\pm$    0.03 &     17.21 $\pm$    0.03 &     15.50 $\pm$    0.00 &     14.62 $\pm$    0.00 &     14.34 $\pm$    0.00 &     14.21 $\pm$    0.03   \\
  Arp 192 &        S &      NGC 3303NED02 &     19.17 $\pm$    0.10 &     18.34 $\pm$    0.03 &     15.94 $\pm$    0.02 &     14.12 $\pm$    0.00 &     13.48 $\pm$    0.00 &     13.03 $\pm$    0.00 &     12.72 $\pm$    0.01   \\
  Arp 192 &         TAIL &      NGC    3303 TAIL &     $\ge$17.85     &     17.82 $\pm$    0.18 &     $\ge$16.78     &     14.55 $\pm$    0.03 &     14.00 $\pm$    0.03 &     13.66 $\pm$    0.02 &     $\ge$13.44       \\
  Arp 202 &        N &      NGC     2719 &     15.88 $\pm$    0.00 &     15.35 $\pm$    0.00 &     14.89 $\pm$    0.01 &     14.10 $\pm$    0.00 &     13.88 $\pm$    0.00 &     13.77 $\pm$    0.01 &     13.72 $\pm$    0.02   \\
  Arp 202 &        S &      NGC    2719A &     15.48 $\pm$    0.00 &     15.16 $\pm$    0.00 &     14.74 $\pm$    0.01 &     14.15 $\pm$    0.00 &     13.85 $\pm$    0.00 &     13.91 $\pm$    0.01 &     13.90 $\pm$    0.03   \\
  Arp 202 &       W TAIL &      ARP 202 WEST TAIL &     21.13 $\pm$    0.14 &     20.67 $\pm$    0.05 &     $\ge$19.81     &     $\ge$20.64     &     $\ge$20.09     &     $\ge$19.05     &     $\ge$18.00       \\
  Arp 202 &         TDG &                   &     19.60 $\pm$    0.03 &     19.61 $\pm$    0.01 &     20.21 $\pm$    0.31 &     20.07 $\pm$    0.12 &     19.86 $\pm$    0.17 &     $\ge$19.36     &     $\ge$18.28       \\
  Arp 242 &         BRIDGE &      NGC    4676 BRIDGE &     20.50 $\pm$    0.05 &     19.66 $\pm$    0.02 &     18.23 $\pm$    0.08 &     16.63 $\pm$    0.01 &     15.84 $\pm$    0.00 &     15.53 $\pm$    0.01 &     15.41 $\pm$    0.02   \\
  Arp 242 &        N &      NGC    4676A &     18.37 $\pm$    0.02 &     17.83 $\pm$    0.01 &     16.40 $\pm$    0.05 &     14.82 $\pm$    0.01 &     14.03 $\pm$    0.00 &     13.62 $\pm$    0.01 &     13.34 $\pm$    0.01   \\
  Arp 242 &       N TAIL &      NGC   4676A TAIL &     18.52 $\pm$    0.03 &     18.06 $\pm$    0.01 &     17.20 $\pm$    0.11 &     15.66 $\pm$    0.01 &     15.23 $\pm$    0.01 &     14.90 $\pm$    0.02 &     14.75 $\pm$    0.04   \\
  Arp 242 &       N TDG &                   &     19.66 $\pm$    0.07 &     19.22 $\pm$    0.02 &     18.37 $\pm$    0.27 &     16.64 $\pm$    0.03 &     16.23 $\pm$    0.01 &     15.82 $\pm$    0.04 &     15.74 $\pm$    0.09   \\
  Arp 242 &        S &      NGC   4676B &     17.49 $\pm$    0.01 &     17.04 $\pm$    0.00 &     16.00 $\pm$    0.04 &     14.39 $\pm$    0.00 &     13.68 $\pm$    0.00 &     13.31 $\pm$    0.00 &     13.04 $\pm$    0.01   \\
  Arp 242 &       S TAIL &      NGC  4676B TAIL &     18.73 $\pm$    0.07 &     18.23 $\pm$    0.02 &     $\ge$17.42     &     15.76 $\pm$    0.03 &     15.22 $\pm$    0.01 &     14.94 $\pm$    0.04 &     14.69 $\pm$    0.08   \\
  Arp 242 &       S TDG &                   &     20.24 $\pm$    0.11 &     19.89 $\pm$    0.04 &     $\ge$18.44     &     18.00 $\pm$    0.08 &     17.28 $\pm$    0.03 &     17.01 $\pm$    0.10 &     16.87 $\pm$    0.23   \\
  Arp 244 &        N &      NGC     4038 &     12.93 $\pm$    0.00 &     12.35 $\pm$    0.00 &        $-$ &        $-$ &        $-$ &        $-$ &        $-$   \\
  Arp 244 &       N TAIL &      NGC    4038 TAIL &     19.10 $\pm$    0.11 &     18.28 $\pm$    0.05 &        $-$ &        $-$ &        $-$ &        $-$ &        $-$   \\
  Arp 244 &        S &      NGC     4039 &     15.19 $\pm$    0.01 &     14.42 $\pm$    0.00 &        $-$ &        $-$ &        $-$ &        $-$ &        $-$   \\
  Arp 244 &       S TAIL &      NGC    4039 TAIL &     17.71 $\pm$    0.11 &     16.67 $\pm$    0.04 &        $-$ &        $-$ &        $-$ &        $-$ &        $-$   \\
  Arp 244 &         TDG &                   &     17.29 $\pm$    0.09 &     16.73 $\pm$    0.05 &        $-$ &        $-$ &        $-$ &        $-$ &        $-$   \\
  Arp 245 &         BRIDGE &      NGC  2992/3 BRIDGE &     17.08 $\pm$    0.11 &     16.45 $\pm$    0.04 &        $-$ &        $-$ &        $-$ &        $-$ &        $-$   \\
  Arp 245 &        N &      NGC     2992 &     16.00 $\pm$    0.03 &     15.32 $\pm$    0.01 &        $-$ &        $-$ &        $-$ &        $-$ &        $-$   \\
  Arp 245 &       N TAIL &      NGC 2992 NORTH TAIL &     17.81 $\pm$    0.16 &     17.11 $\pm$    0.05 &        $-$ &        $-$ &        $-$ &        $-$ &        $-$   \\
  Arp 245 &        S &      NGC     2993 &     13.51 $\pm$    0.01 &     13.10 $\pm$    0.00 &        $-$ &        $-$ &        $-$ &        $-$ &        $-$   \\
  Arp 245 &      SE TAIL &      NGC 2993 SOUTHEAST TAIL &     16.98 $\pm$    0.16 &     16.48 $\pm$    0.06 &        $-$ &        $-$ &        $-$ &        $-$ &        $-$   \\
  Arp 245 &         TDG &                   &     17.68 $\pm$    0.04 &     17.19 $\pm$    0.01 &        $-$ &        $-$ &        $-$ &        $-$ &        $-$   \\
  Arp 253 &        E &     UGCA      174 &     17.07 $\pm$    0.01 &     16.74 $\pm$    0.00 &        $-$ &        $-$ &        $-$ &        $-$ &        $-$   \\
  Arp 253 &        W &     UGCA      173 &     15.95 $\pm$    0.00 &     15.72 $\pm$    0.00 &        $-$ &        $-$ &        $-$ &        $-$ &        $-$   \\
  Arp 254 &         BRIDGE &      ARP     254 BRIDGE &     $\ge$17.80     &     17.64 $\pm$    0.12 &        $-$ &        $-$ &        $-$ &        $-$ &        $-$   \\
  Arp 254 &        N &      NGC     5917 &     14.47 $\pm$    0.01 &     14.15 $\pm$    0.01 &        $-$ &        $-$ &        $-$ &        $-$ &        $-$   \\
  Arp 254 &        S &      MCG -01-39-003 &     16.24 $\pm$    0.03 &     15.72 $\pm$    0.01 &        $-$ &        $-$ &        $-$ &        $-$ &        $-$   \\
  Arp 254 &       S TAIL &      MCG -01-39-003 SOUTH TAIL &     19.01 $\pm$    0.12 &     18.30 $\pm$    0.03 &        $-$ &        $-$ &        $-$ &        $-$ &        $-$   \\
  Arp 261 &        N &      Arp 261NED04 &     16.35 $\pm$    0.01 &     16.05 $\pm$    0.00 &        $-$ &        $-$ &        $-$ &        $-$ &        $-$   \\
  Arp 261 &  N SMALL &      KTS      52C &     17.17 $\pm$    0.01 &     16.77 $\pm$    0.00 &        $-$ &        $-$ &        $-$ &        $-$ &        $-$   \\
  Arp 261 &       N TAIL &      Arp 261 NORTHWEST TAIL &     17.17 $\pm$    0.02 &     16.88 $\pm$    0.01 &        $-$ &        $-$ &        $-$ &        $-$ &        $-$   \\
  Arp 261 &        S &      Arp 261NED01 &     15.19 $\pm$    0.00 &     14.91 $\pm$    0.00 &        $-$ &        $-$ &        $-$ &        $-$ &        $-$   \\
  Arp 261 &      SW TAIL &                   &     18.83 $\pm$    0.04 &     18.39 $\pm$    0.02 &        $-$ &        $-$ &        $-$ &        $-$ &        $-$   \\
  Arp 269 &        N &      NGC     4485 &     13.66 $\pm$    0.00 &     13.32 $\pm$    0.00 &     12.96 $\pm$    0.01 &     12.03 $\pm$    0.01 &     11.78 $\pm$    0.01 &     11.65 $\pm$    0.01 &     11.60 $\pm$    0.04   \\
  Arp 269 &        S &      NGC     4490 &     17.18 $\pm$    0.01 &     16.93 $\pm$    0.01 &     16.89 $\pm$    0.09 &     15.78 $\pm$    0.03 &     15.34 $\pm$    0.04 &     15.16 $\pm$    0.05 &     15.05 $\pm$    0.14   \\
  Arp 270 &        E &      NGC     3396 &     14.19 $\pm$    0.00 &     13.85 $\pm$    0.00 &     13.39 $\pm$    0.03 &     12.33 $\pm$    0.04 &     12.07 $\pm$    0.05 &     11.90 $\pm$    0.08 &     11.85 $\pm$    0.11   \\
  Arp 270 &        3$'$ from TIP &                   &     20.39 $\pm$    0.08 &     20.15 $\pm$    0.06 &     $\ge$18.96     &     $\ge$18.63     &     $\ge$19.30     &     $\ge$17.16     &     $\ge$17.41       \\
  Arp 270 &       S TAIL &                   &     $\ge$19.89     &     19.46 $\pm$    0.19 &     $\ge$17.02     &     $\ge$15.83     &     $\ge$15.08     &     $\ge$14.49     &     $\ge$14.10       \\
  Arp 270 &        W &      NGC     3395 &     13.46 $\pm$    0.00 &     13.16 $\pm$    0.00 &     12.90 $\pm$    0.01 &     12.06 $\pm$    0.02 &     11.77 $\pm$    0.03 &     11.63 $\pm$    0.05 &     11.63 $\pm$    0.07   \\
  Arp 270 &         TDG &                   &     $\ge$20.70     &     20.48 $\pm$    0.22 &     $\ge$17.89     &     17.42 $\pm$    0.30 &     16.78 $\pm$    0.08 &     $\ge$15.83     &     $\ge$16.32       \\
  Arp 271 &        N &      NGC     5427 &     13.71 $\pm$    0.00 &     13.19 $\pm$    0.00 &        $-$ &        $-$ &        $-$ &        $-$ &        $-$   \\
  Arp 271 &        S &      NGC     5426 &     14.52 $\pm$    0.00 &     14.13 $\pm$    0.00 &        $-$ &        $-$ &        $-$ &        $-$ &        $-$   \\
  Arp 280 &        E &      NGC    3769A &     16.31 $\pm$    0.00 &     15.95 $\pm$    0.00 &     15.58 $\pm$    0.04 &     14.74 $\pm$    0.01 &     14.45 $\pm$    0.01 &     14.29 $\pm$    0.01 &     14.17 $\pm$    0.09   \\
  Arp 280 &        W &      NGC     3769 &     14.01 $\pm$    0.00 &     13.55 $\pm$    0.00 &     12.40 $\pm$    0.01 &     11.20 $\pm$    0.00 &     10.80 $\pm$    0.00 &     10.46 $\pm$    0.00 &     10.26 $\pm$    0.01   \\
  Arp 282 &       E TAIL &      NGC     169 TAIL &     $\ge$19.49     &     19.39 $\pm$    0.09 &     $\ge$17.17     &     15.53 $\pm$    0.30 &     $\ge$14.70     &     $\ge$14.21     &     14.85 $\pm$    0.36   \\
  Arp 282 &        N &      NGC      169 &     18.47 $\pm$    0.12 &     17.80 $\pm$    0.02 &     15.53 $\pm$    0.07 &     13.65 $\pm$    0.06 &     12.75 $\pm$    0.05 &     12.19 $\pm$    0.05 &     11.80 $\pm$    0.02   \\
  Arp 282 &        S &      NGC     169A &     17.35 $\pm$    0.01 &     16.90 $\pm$    0.00 &     15.75 $\pm$    0.03 &     14.50 $\pm$    0.04 &     13.89 $\pm$    0.05 &     13.52 $\pm$    0.05 &     13.32 $\pm$    0.03   \\
  Arp 283 &         BRIDGE &      NGC  2798/9 BRIDGE &     18.98 $\pm$    0.04 &     18.52 $\pm$    0.01 &     18.60 $\pm$    0.25 &     16.91 $\pm$    0.02 &     16.49 $\pm$    0.03 &     16.35 $\pm$    0.04 &     16.73 $\pm$    0.16   \\
  Arp 283 &        E &      NGC     2799 &     16.22 $\pm$    0.01 &     15.87 $\pm$    0.00 &     15.09 $\pm$    0.02 &     13.98 $\pm$    0.00 &     13.48 $\pm$    0.01 &     13.26 $\pm$    0.01 &     13.17 $\pm$    0.02   \\
  Arp 283 &       N TAIL &      NGC 2798 NORTH TAIL &     $\ge$20.48     &     19.23 $\pm$    0.03 &     17.03 $\pm$    0.11 &     15.15 $\pm$    0.01 &     14.59 $\pm$    0.01 &     14.25 $\pm$    0.01 &     14.14 $\pm$    0.03   \\
  Arp 283 &       S TAIL &      NGC 2798 SOUTH TAIL &     20.53 $\pm$    0.31 &     19.52 $\pm$    0.05 &     $\ge$18.02     &     16.44 $\pm$    0.02 &     15.82 $\pm$    0.04 &     15.47 $\pm$    0.03 &     15.53 $\pm$    0.10   \\
  Arp 283 &        W &      NGC     2798 &     16.08 $\pm$    0.01 &     15.26 $\pm$    0.00 &     13.85 $\pm$    0.01 &     12.69 $\pm$    0.00 &     11.86 $\pm$    0.00 &     11.56 $\pm$    0.00 &     11.34 $\pm$    0.00   \\
Arp 284 & BRIDGE & NGC 7714/5 BRIDGE & 17.34 $\pm$ 0.01 & $-$     & $-$ & $-$ & $-$ & $-$ & $-$ \\
Arp 284 & E & NGC 7715 & 17.10 $\pm$ 0.00 & $-$     & $-$ & $-$ & $-$ & $-$ & $-$ \\
Arp 284 & E TAIL & NGC 7715 TAIL & 17.45 $\pm$ 0.01 & $-$     & $-$ & $-$ & $-$ & $-$ & $-$ \\
Arp 284 & W & NGC 7714 & 14.37 $\pm$ 0.00 & $-$     & $-$ & $-$ & $-$ & $-$ & $-$ \\
Arp 284 & W TAIL & NGC 7714 TAIL & 15.18 $\pm$ 0.00 & $-$     & $-$ & $-$ & $-$ & $-$ & $-$ \\
  Arp 285 &         BRIDGE &      NGC  2854/6 BRIDGE &        $-$ &     20.46 $\pm$    0.15 &     $\ge$18.31     &     17.99 $\pm$    0.16 &     16.44 $\pm$    0.07 &     15.58 $\pm$    0.07 &     15.30 $\pm$    0.06   \\
  Arp 285 &        N &      NGC     2856 &        $-$ &     15.97 $\pm$    0.01 &     14.47 $\pm$    0.02 &     13.03 $\pm$    0.00 &     12.30 $\pm$    0.00 &     11.98 $\pm$    0.01 &     11.79 $\pm$    0.00   \\
  Arp 285 &       N TAIL &      NGC    2856 TAIL &        $-$ &     19.37 $\pm$    0.02 &     19.44 $\pm$    0.28 &     18.00 $\pm$    0.04 &     17.93 $\pm$    0.07 &     17.49 $\pm$    0.15 &     17.98 $\pm$    0.19   \\
  Arp 285 &        S &      NGC     2854 &        $-$ &     15.80 $\pm$    0.01 &     14.75 $\pm$    0.03 &     13.41 $\pm$    0.01 &     12.82 $\pm$    0.01 &     12.49 $\pm$    0.01 &     12.33 $\pm$    0.01   \\
  Arp 285 &       S TAIL &      NGC 2854 SOUTH TAIL &        $-$ &     20.12 $\pm$    0.13 &     $\ge$18.21     &     $\ge$18.29     &     $\ge$17.48     &     $\ge$17.14     &     $\ge$16.53       \\
  Arp 290 &        N &       IC      196 &     17.03 $\pm$    0.03 &     16.41 $\pm$    0.02 &     15.25 $\pm$    0.07 &     13.33 $\pm$    0.01 &     12.54 $\pm$    0.01 &     12.13 $\pm$    0.01 &     11.95 $\pm$    0.03   \\
  Arp 290 &       N TAIL &       IC     196 TAIL &     17.89 $\pm$    0.02 &     17.29 $\pm$    0.01 &     17.18 $\pm$    0.07 &     15.42 $\pm$    0.01 &     14.84 $\pm$    0.01 &     14.44 $\pm$    0.01 &     14.32 $\pm$    0.04   \\
  Arp 290 &        S &       IC      195 &     20.42 $\pm$    0.23 &     18.59 $\pm$    0.06 &     15.78 $\pm$    0.05 &     13.81 $\pm$    0.01 &     13.04 $\pm$    0.01 &     12.62 $\pm$    0.01 &     12.42 $\pm$    0.02   \\
  Arp 295 &         BRIDGE &      ARP     295 BRIDGE &     $\ge$20.19     &     19.12 $\pm$    0.14 &        $-$ &        $-$ &        $-$ &        $-$ &        $-$   \\
  Arp 295 &        N &      ARP    295B &     16.75 $\pm$    0.01 &     16.21 $\pm$    0.01 &        $-$ &        $-$ &        $-$ &        $-$ &        $-$   \\
  Arp 295 &        S &      ARP     295A &     19.75 $\pm$    0.13 &     18.34 $\pm$    0.05 &        $-$ &        $-$ &        $-$ &        $-$ &        $-$   \\
  Arp 295 &       S TAIL &      ARP    295A TAIL &     20.04 $\pm$    0.17 &     19.06 $\pm$    0.09 &        $-$ &        $-$ &        $-$ &        $-$ &        $-$   \\
  Arp 297 &       NE &      NGC     5755 &     17.48 $\pm$    0.01 &     16.73 $\pm$    0.01 &     15.95 $\pm$    0.05 &     14.43 $\pm$    0.01 &     13.99 $\pm$    0.01 &     13.75 $\pm$    0.01 &     13.66 $\pm$    0.05   \\
  Arp 297 &    NE N TAIL &      NGC    5755 TAIL &     18.86 $\pm$    0.04 &     18.08 $\pm$    0.02 &     $\ge$18.09     &     16.62 $\pm$    0.06 &     16.37 $\pm$    0.04 &     16.22 $\pm$    0.08 &     $\ge$15.87       \\
  Arp 297 &       NW &      NGC     5753 &     18.74 $\pm$    0.03 &     18.30 $\pm$    0.02 &     17.56 $\pm$    0.13 &     15.69 $\pm$    0.02 &     15.03 $\pm$    0.01 &     14.69 $\pm$    0.02 &     14.62 $\pm$    0.07   \\
  Arp 297 &       SE &      NGC     5754 &     15.96 $\pm$    0.01 &     15.61 $\pm$    0.01 &     14.86 $\pm$    0.07 &     13.19 $\pm$    0.01 &     12.57 $\pm$    0.01 &     12.26 $\pm$    0.01 &     12.14 $\pm$    0.05   \\
  Arp 297 &      SE TAIL &      NGC    5754 TAIL &     20.07 $\pm$    0.07 &     19.35 $\pm$    0.04 &     $\ge$18.62     &     17.13 $\pm$    0.06 &     16.66 $\pm$    0.03 &     16.21 $\pm$    0.05 &     $\ge$16.48       \\
  Arp 297 &       SW &      NGC     5752 &     18.17 $\pm$    0.02 &     17.56 $\pm$    0.01 &     16.40 $\pm$    0.04 &     15.03 $\pm$    0.01 &     14.44 $\pm$    0.01 &     14.16 $\pm$    0.01 &     14.01 $\pm$    0.04   \\
  Arp 297 &       W TAIL &      NGC    5452 TAIL &     $\ge$20.45     &     19.46 $\pm$    0.12 &     $\ge$17.56     &     17.13 $\pm$    0.16 &     16.55 $\pm$    0.07 &     16.12 $\pm$    0.10 &     $\ge$15.47       \\
  Arp 298 &        N &       IC     5283 &     18.20 $\pm$    0.16 &     17.36 $\pm$    0.07 &        $-$ &        $-$ &        $-$ &        $-$ &        $-$   \\
  Arp 298 &        S &      NGC     7469 &     14.57 $\pm$    0.01 &     14.04 $\pm$    0.01 &        $-$ &        $-$ &        $-$ &        $-$ &        $-$   \\
  Arp 298 &       W TAIL &       IC    5283 TAIL &     $\ge$19.37     &     $\ge$19.34     &        $-$ &        $-$ &        $-$ &        $-$ &        $-$   \\
  Arp 305 &         BRIDGE &      NGC  4016/7 BRIDGE &     16.89 $\pm$    0.01 &     16.61 $\pm$    0.00 &     16.45 $\pm$    0.05 &     15.37 $\pm$    0.02 &     15.10 $\pm$    0.04 &     15.07 $\pm$    0.10 &     15.09 $\pm$    0.10   \\
  Arp 305 &        N &      NGC     4016 &     15.40 $\pm$    0.01 &     15.19 $\pm$    0.00 &     15.02 $\pm$    0.03 &     14.12 $\pm$    0.01 &     13.80 $\pm$    0.02 &     13.58 $\pm$    0.06 &     13.56 $\pm$    0.05   \\
  Arp 305 &       N TAIL &      NGC 4016 NORTH TAIL &     17.51 $\pm$    0.03 &     17.27 $\pm$    0.01 &     18.03 $\pm$    0.27 &     16.47 $\pm$    0.07 &     16.09 $\pm$    0.11 &     15.85 $\pm$    0.25 &     $\ge$16.00       \\
  Arp 305 &        S &      NGC     4017 &     15.44 $\pm$    0.01 &     15.05 $\pm$    0.00 &     14.50 $\pm$    0.02 &     13.44 $\pm$    0.01 &     13.00 $\pm$    0.01 &     12.83 $\pm$    0.03 &     12.72 $\pm$    0.02   \\
  Arp 305 &       S TAIL &      NGC 4017 SOUTH TAIL &     16.44 $\pm$    0.07 &     16.14 $\pm$    0.02 &     $\ge$15.94     &     14.89 $\pm$    0.12 &     14.34 $\pm$    0.16 &     $\ge$13.78     &     $\ge$13.90       \\
  Arp 305 &         TDG &                   &     18.16 $\pm$    0.02 &     18.02 $\pm$    0.01 &     $\ge$18.73     &     17.92 $\pm$    0.13 &     17.79 $\pm$    0.27 &     $\ge$16.71     &     $\ge$16.68       \\
  NGC 4567 &        N &                   &        $-$ &     14.13 $\pm$    0.01 &     12.80 $\pm$    0.02 &     11.44 $\pm$    0.01 &     10.78 $\pm$    0.01 &     10.47 $\pm$    0.01 &     10.29 $\pm$    0.02   \\
  NGC 4567 &        S &                   &        $-$ &     13.69 $\pm$    0.01 &     12.61 $\pm$    0.02 &     11.15 $\pm$    0.01 &     10.52 $\pm$    0.01 &     10.14 $\pm$    0.01 &      9.89 $\pm$    0.02   \\
\enddata
\end{deluxetable}
\end{landscape}

\clearpage

\clearpage

\clearpage
\thispagestyle{empty}
\clearpage




\end{document}